\documentclass[conference]{IEEEtran}
\IEEEoverridecommandlockouts
\usepackage{hyperref}
\usepackage{amsmath,amssymb,amsfonts}
\usepackage{algorithmic}
\usepackage{graphicx}
\usepackage{bbm}
\usepackage{amsmath}
\usepackage{textcomp}
\usepackage{xcolor}
\usepackage{quantikz}
\usepackage{pgfplots}
\usepgfplotslibrary{groupplots}
\usepackage{placeins}
\usepackage{caption}
\usepackage{subcaption}
\usepackage{multirow}
\usepackage{booktabs}
\usepackage{flushend}
\def\BibTeX{{\rm B\kern-.05em{\sc i\kern-.025em b}\kern-.08em
    T\kern-.1667em\lower.7ex\hbox{E}\kern-.125emX}}

\DeclareCaptionLabelSeparator{periodspace}{.\quad}

\begin{document}
\bstctlcite{IEEEexample:BSTcontrol}

\title{On Modifying the Variational Quantum Singular
Value Decomposition Algorithm\\
}

\author{\IEEEauthorblockN{Jezer Jojo\IEEEauthorrefmark{1}\IEEEauthorrefmark{2}, Ankit Khandelwal\IEEEauthorrefmark{2} and M Girish Chandra\IEEEauthorrefmark{2}}
\IEEEauthorblockA{\IEEEauthorrefmark{1}Indian Institute of Science Education and Research, Pune, India}
\IEEEauthorblockA{\IEEEauthorrefmark{2}TCS Research, Tata Consultancy Services, India\\
Email: \href{mailto:jezer.vallivattam@students.iiserpune.ac.in}{jezer.vallivattam@students.iiserpune.ac.in}, \href{mailto:khandelwal.ankit3@tcs.com}{khandelwal.ankit3@tcs.com}, \href{mailto:m.gchandra@tcs.com}{m.gchandra@tcs.com}}}

\maketitle
\begin{abstract}
In this work, we discuss two modifications that can be made to a known variational quantum singular value decomposition algorithm popular in the literature. The first is a change to the objective function which hints at improved performance of the algorithm. The second modification introduces a new way of computing expectation values of general matrices, which is a key step in the algorithm. We then benchmark this modified algorithm and compare the performance of our new objective function with the existing one.

\end{abstract}

\section{Introduction}
The Singular Value Decomposition (SVD) is a kind of matrix decomposition that finds many uses in a variety of fields. Its use is ubiquitous in the field of Data Science where the SVD is often used to reduce the dimensionality of input data \cite{dimred}. This specific use of SVD finds itself in applications like Principal Component Analysis (PCA) and recommendation systems via matrix completion \cite{rec}. Finding the truncated SVD of an image is also known to serve as a method of image compression \cite{svdimgcomp}.\\
Quantum computing is a new paradigm of computing that makes use of the principles of quantum mechanics \cite{feynman}. It is known to offer advantage over classical computers in certain problems such as unstructured search \cite{grover} and factoring primes \cite{shor}.\\
The inherent linearity of quantum mechanics suggests the existence of efficient quantum algorithms to solve problems in linear algebra like computing the SVD of a matrix. In fact, purely quantum algorithms for finding a matrix's SVD do exist \cite{kp,bellante} but require large circuits with too many qubits to be feasibly implemented in the near term.\\
In recent work \cite{wang} by Xin Wang et al., a hybrid variational algorithm that computes the SVD of a given square matrix was proposed. This algorithm uses both classical and quantum computers and requires less qubits than purely quantum approaches. Another variational algorithm to compute the SVD has been proposed in \cite{qsvdecomposer}.\\
In this work, we discuss two modifications that can be made independently to this variational quantum singular value decomposition algorithm. We start by defining our notation in Section \ref{sec:notation}. Then we give a brief overview of the existing algorithm in Section \ref{sec:overview}. The first modification we apply is a change to the objective function which we discuss in Section \ref{sec:obj}. This new objective function makes stronger theoretical guarantees than the existing one. The second modification is discussed in Section \ref{sec:exp} and it introduces a possibly new way of computing expectation values, which is a key step in the algorithm. This approach could prove useful in cases where we only have quantum data of the matrix we wish to find the SVD of. However, it is worth mentioning that the variational algorithm proposed in \cite{qsvdecomposer} also takes the matrix as a quantum input; but it has certain limitations compared to our method in that it does not provide a means to retrieve the relative phases between singular vectors.

Beyond the context of computing SVD, our new method serves as a new block encoding construction. While its subnormalization factor is higher than most other methods \cite{fable}\cite{blockencoding}, it is unique - to the best of our knowledge - in that it takes the input matrix directly as an amplitude encoded statevector and doesn't require classical knowledge of the matrix or any kind of oracle access to the matrix.

In Section \ref{sec:imp} we describe our overall modified algorithm and the details of our implementation. In Section \ref{sec:bench}, we then benchmark this modified algorithm and compare the performance of the modified objective function against the original one. Finally, we discuss the implications of our results and future works in Section \ref{sec:discuss}


\begin{figure}[tbp]
    \centering
    \begin{subfigure}{0.3\columnwidth}
        \centering
        \includegraphics[width=2.5cm]{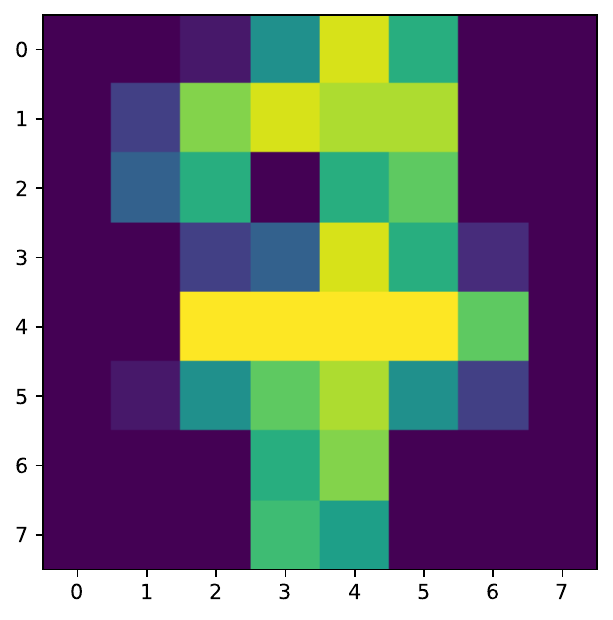}
        \caption{Original Image \\ \hphantom{.}}
    \end{subfigure}
    \begin{subfigure}{0.3\columnwidth}
        \centering
        \includegraphics[width=2.5cm]{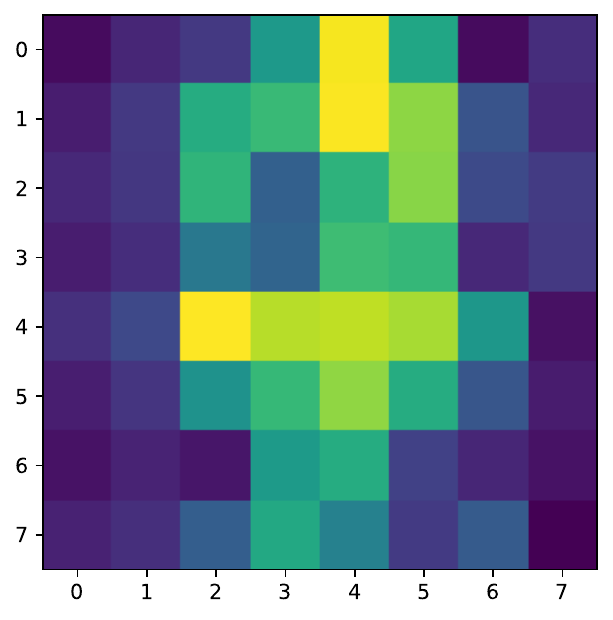}
        \caption{$f'_8=3.4840$\\MSE $=3.9566$}
    \end{subfigure}
    \begin{subfigure}{0.3\columnwidth}
        \centering
        \includegraphics[width=2.5cm]{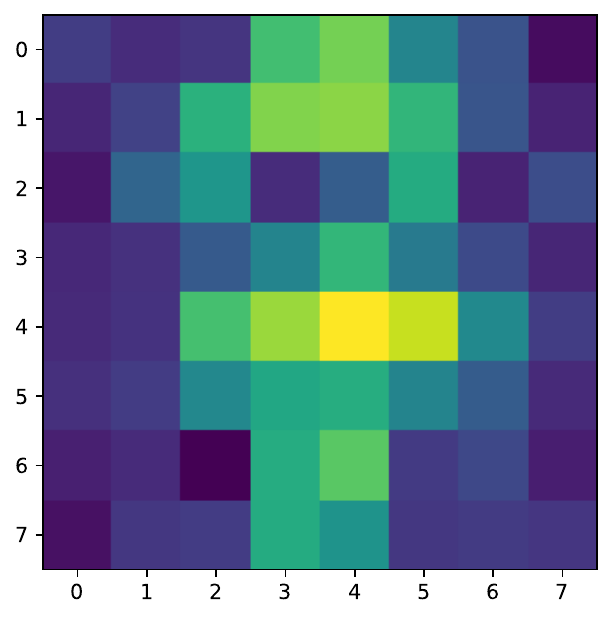}
        \caption{$f'_8=3.6658$\\MSE $=6.0184$}
    \end{subfigure}
    \caption{The first image (a) is taken from the MNIST handwritten digits dataset \cite{mnist}. Images (b) and (c) are reconstructions obtained by running the algorithm for $T=3$ and $T=8$ respectively using the original objective function with a 4-layer ansatz of the form in Fig. \ref{fig:ansatz}. $f'_8$ is calculated for both of them. As we can see, the $T=8$ optimization picks image (c) over image (b) because of the higher $f'_8$ value of (c) even though the quality of (b) is better. This explains the results in Fig. \ref{fig:rand_mse}. }%
    \label{fig:orig_fun_eg}%
    \vspace{-5mm}
\end{figure}

\section{Notation}\label{sec:notation}
\begin{itemize}
    \item Let $A$ be a $2^n\times2^n$ square matrix that we want the SVD of.
    \item The SVD of $A$ is given by $A=U\Sigma V^\dagger$ where $U$ and $V$ are unitary matrices and $\Sigma$ is a diagonal matrix of positive real values arranged in descending order.\\
    \item Let $\sigma_i=\Sigma_{ii}$ denote the $i$-th largest singular value of $A$.
    \item Just as $A=\sum_i^{2^n} \sigma_i U|i\rangle\langle i|V^\dagger$, we can define the truncated matrix, $A_{trunc}^T:=\sum_i^T \sigma_i U|i\rangle\langle i|V^\dagger$.
    \item We define the following matrices based on the Pauli basis matrices:
    \begin{align*}
        &P_0=\begin{bmatrix}
                1 & 0\\
                0 & 1
                \end{bmatrix}, 
        &&P_1=\begin{bmatrix}
                0 & 1\\
                1 & 0
                \end{bmatrix},\\
        &P_2=\begin{bmatrix}
                0 & 1\\
                -1 & 0
                \end{bmatrix}, 
        &&P_3=\begin{bmatrix}
                1 & 0\\
                0 & -1
                \end{bmatrix}
    \end{align*}
    \item We define $P_{s_1,\dots,s_n}:=P_{s_1}\otimes\dots\otimes P_{s_n}$.
    \item There exists a unique set $\{c_{s_1,\dots,s_n}\}$ such that $A=\sum_{s_1,\dots,s_n}c_{s_1,\dots,s_n}P_{s_1,\dots,s_n}$. Or, for convenience
    \begin{equation}
    A=\sum_{\Vec{s}}c_{\Vec{s}}P_{\Vec{s}}
    \end{equation}
\end{itemize}

\section{Overview of the algorithm}\label{sec:overview}
In this section, we describe the general strategy behind the existing algorithm proposed in \cite{wang}. We lay out this description in a way that makes it convenient for us to frame our modifications later.

The algorithm takes the following inputs
\begin{enumerate}
    \item A $2^n\times2^n$ matrix $A$
    \item A positive integer $1\leq T\leq 2^n$
\end{enumerate}

If the matrix we want the SVD of is not $2^n\times2^n$, we can always pad it with $0$'s to make it so. 

It then returns quantum circuits $\Tilde{U}$ and $\Tilde{V}$ along with a sequence of numbers $\Tilde{\sigma}_1\dots\Tilde{\sigma}_T$ such that the following holds:

\begin{equation}\label{approx}
    A^T_{rec}:=\sum_i^T \Tilde{\sigma}_i\Tilde{U}|i\rangle\langle i|\Tilde{V}^\dagger \approx \sum_i^T \sigma_i U|i\rangle\langle i|V^\dagger
\end{equation}

The way the algorithm does this is by having an ansatz $W$ and two sets of parameters $\Vec{\alpha}$ and $\Vec{\beta}$ which are optimized using some objective function (which we talk about in the next section). This objective function is crafted such that for optimized parameters $\Vec{\alpha}*$ and $\Vec{\beta}*$, $W(\Vec{\alpha}*)=\Tilde{U}$ and $W(\Vec{\beta}*)=\Tilde{V}$. The quality of the approximation in \eqref{approx} is dependent on how well this optimization process goes. To then get $\Tilde{\sigma}_i$, we evaluate $\Tilde{\sigma}_i=\langle i|\Tilde{U}^\dagger A\Tilde{V}|i\rangle$.

\section{New objective function}\label{sec:obj}

The first modification we make to the algorithm in \cite{wang} is a change in the objective function. The objective function used there by Xin Wang et al. was

\begin{equation}
    f'_T(\Vec{\alpha},\Vec{\beta})=\sum_i^T q_i \text{Re}(\langle i| W(\Vec{\alpha})^\dagger A W(\Vec{\beta})|i\rangle)
\end{equation}
where $q_i\in\mathbbm{R}^+$ and $i<j\implies q_j<q_i$ and any such choice for $\{q_i\}$ works.

They then proved that the theoretical maximum of this objective function corresponded to $W(\Vec{\alpha})=U$ and $W(\Vec{\beta})=V$. In practice however we usually don't find the true theoretical maximum; instead we use gradient descent to find one of possibly many local maxima with $W$ often constrained to a limited subspace of all possible unitaries. What we would ideally want from an objective function then is some guarantee that a higher value of the objective function corresponds to a better solution. This however is not true for the proposed objective function as we can see in Fig. \ref{fig:orig_fun_eg}.

To address this issue, we propose a new objective function instead:

\begin{equation}
    f_T(\Vec{\alpha},\Vec{\beta})=\sum_i^T |\langle i| W(\Vec{\alpha})^\dagger A W(\Vec{\beta})|i\rangle|^2
\end{equation}

The advantage behind using this objective function is that a larger $f_T$ corresponds to a better SVD solution. Of course we need to pick a metric to explain what we mean by ``a better SVD solution''. The metric we choose for our purpose is the Mean Square Error (MSE) of the entries of our reconstructed matrix $A^T_{rec}$ with respect to $A$; i.e. $\frac{1}{2^{2n}}||A-A^T_{rec}||_F^2$, where $||.||_F$ denotes the Frobenius norm.\\
So now, the claim we make is formalized below:

\begin{align}
   & f_T(\Vec{\alpha}_1,\Vec{\beta}_1)>f_T(\Vec{\alpha}_2,\Vec{\beta}_2)\iff\nonumber\\
  &\text{MSE of }A^T_{rec}(\Vec{\alpha}_1,\Vec{\beta}_1)<\text{MSE of }A^T_{rec}(\Vec{\alpha}_2,\Vec{\beta}_2)\label{thm}
\end{align}

The proof is as follows

\begin{align}
& \text{MSE of }A^T_{rec}(\Vec{\alpha},\Vec{\beta})\nonumber\\
=&\frac{1}{2^{2n}}||A-A^T_{rec}(\Vec{\alpha},\Vec{\beta})||_F^2\nonumber\\
=&\frac{1}{2^{2n}}||W^\dagger(\Vec{\alpha})(A-A^T_{rec}(\Vec{\alpha},\Vec{\beta}))W(\Vec{\beta})||_F^2\nonumber\\
=&\frac{1}{2^{2n}}\sum_{l,m}|\langle l|W^\dagger(\Vec{\alpha})(A-A^T_{rec}(\Vec{\alpha},\Vec{\beta}))W(\Vec{\beta})|m\rangle|^2\label{eq_rec}\\
=&\frac{1}{2^{2n}}\left(\sum_{l\neq m}|\langle l|W^\dagger(\Vec{\alpha})AW(\Vec{\beta})|m\rangle|^2 +\right.\nonumber\\
&\left.\sum_{l\geq T}|\langle l|W^\dagger(\Vec{\alpha})AW(\Vec{\beta})|l\rangle|^2\right)\label{eq_wo_req}\\
=&\frac{1}{2^{2n}}(||A||_F^2 - f_T(\Vec{\alpha},\Vec{\beta}))\label{final_exp}
\end{align}

To get from \eqref{eq_rec} to \eqref{eq_wo_req}, we use the fact that $A^T_{rec}=\sum_i^T (\langle i|W^\dagger(\Vec{\alpha})AW(\Vec{\beta})|i\rangle)W(\Vec{\alpha})|i\rangle\langle i| W^\dagger(\Vec{\beta})$.\\
Since $||A||_F$ is constant in $(\Vec{\alpha},\Vec{\beta})$, we see in \eqref{final_exp} that any increase in $f_T$ is directly associated with a corresponding decrease in the MSE and vice versa.

A corollary of \eqref{thm} when combined with the Eckart–Young–Mirsky theorem \cite{eckart} is that at the theoretical optimum of $f_T$, $A^T_{rec}=A_{trunc}^T$.

\section{Novel approach to find the expectation value of a non-unitary matrix}\label{sec:exp}
\subsection{Known method for evaluating the expectation value of a non-unitary operator}

So now we have to figure out how we can compute this objective function for a given input of $\Vec{\alpha}$ and $\Vec{\beta}$. Each term in this expression is the expectation value of the operator $W^\dagger(\Vec{\alpha}) A W(\Vec{\beta})$ for a basis state $|i\rangle$. We have operators $W(\Vec{\alpha})$ and $W(\Vec{\beta})$ that we can run on a quantum computer, but we can't have an operator for $A$ since it is not unitary. The usual way to deal with this \cite{peruzzo,kohda} is to
first find the decomposition of $A$ as the sum of Pauli basis matrix strings, then find expectation values for each of those Pauli strings, and then finally use linearity of expectations to stitch back together the expectation value of $A$. This is the method used in \cite{wang}. We now elaborate on each step of this technique.\\

\textbf{Step 1:} Get the Pauli decomposition of $A$.

\begin{equation}
A=\sum_{\Vec{s}}c_{\Vec{s}}P_{\Vec{s}}
\end{equation}

\textbf{Step 2:} Compute $\langle i| W^\dagger(\Vec{\alpha}) P_{\Vec{s}} W(\Vec{\beta})|i\rangle$ on a quantum computer for all $\Vec{s}$ and for all $i$. This can be done using Hadamard tests to calculate the real and imaginary parts of each expectation value.

\textbf{Step 3:} In this step we express the objective function in terms of the expectation values we computed in step 2 by plugging the Pauli decomposition of $A$ into $f_T$:

\begin{align}
f_T(\Vec{\alpha},\Vec{\beta})&=\sum^T_i |\langle i| W^\dagger(\Vec{\alpha}) \sum_{\Vec{s}}c_{\Vec{s}}P_{\Vec{s}} W(\Vec{\beta})|i\rangle|^2\nonumber\\
&=\sum^T_i |\left(\sum_{\Vec{s}}c_{\Vec{s}}\langle i| W^\dagger(\Vec{\alpha}) P_{\Vec{s}} W(\Vec{\beta})|i\rangle\right)|^2\label{paulicost}
\end{align}

Having used our quantum circuits to compute each expectation value in \eqref{paulicost}, we plug them into this expression to evaluate the final value of our objective function.\\

\subsection{Novel method}

In this section, we discuss our approach for computing the expectation value of a known matrix $M$. Ultimately this method takes an amplitude encoding of $M$ as input and uses it to implement a block encoding of $M$. This method of block encoding has not been proposed before in the literature to the best of our knowledge.

\subsubsection{Circuit registers}
Say $M$ is a square matrix with $2^n$ rows and columns and we want to build a quantum circuit to compute $\langle\psi|M|\psi\rangle$.\\
Our circuit will have a data register of $2n$ qubits denoted $dat$ plus one auxiliary qubit denoted $aux$ and we will use them to store the entries of $M$ in amplitude encoding. We will also have an input state register of $n$ qubits denoted $isr$ that will be used to store the input state $|\psi\rangle$.

\begin{figure}[tb]
   \centering
 \includegraphics[width=0.7\columnwidth]{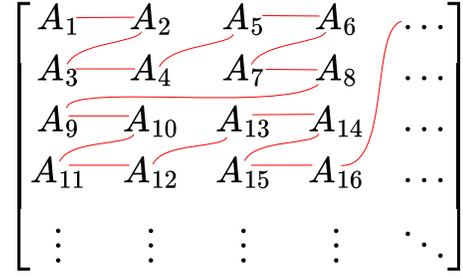}
\caption{Ordering of $A_{flat}$}
\label{fig:zorder}
\end{figure}

\subsubsection{Encoding our matrix as a statevector}
The first thing we need to do is to scale our matrix to ensure that none of the matrix elements are greater than $1$. Next we need to reshape our matrix into a one dimensional vector. The intuitive way to flatten a matrix $M$ into vector $M_{flat}$ for amplitude encoding is by ordering the matrix elements row by row. The problem with this for our purposes is that this ordering does not preserve tensor products. So instead we order our matrix elements using a Z-order space filling curve as shown in Fig. \ref{fig:zorder}.\\

Now, if $X\otimes Y=Z$, then $X_{flat}\otimes Y_{flat}=Z_{flat}$. 
To prove this, let $G$ map square matrices to their Z-order flattened vectors and let $l$ be a function that maps $(i,j)$ to $k$ such that
\begin{equation*}    
\text{revbin}(k)[p]=
\begin{cases} 
  \text{revbin}(i)[p//2] &  \text{if $p$ is odd} \\
  \text{revbin}(j)[p//2] &  \text{if $p$ is even}
\end{cases}
\end{equation*}
Here, $\text{revbin}(x)[p]$ is the $p$-th element (starting from $0$) of the reversed binary bitstring representing $x$ and $//$ represents floored division. Essentially, $l$ takes two numbers and interleaves the bits in their binary representations to generate a new number. Given how Z-ordering is defined, it can be seen that $(G(A))_{l(i,j)}=A_{ij}$
(Note that this means that a row-by-row amplitude encoding of the matrix can be converted to a Z-ordered amplitude encoding by just reordering the qubits).
What we need to prove now is that $(G(A\otimes B))_m=(G(A)\otimes G(B))_m$ for all $m$. Let $(i,j)=l^{-1}(m)$ and let the dimensions of $B$ be $2^q \times 2^q$.
\begin{align}
    LHS&=(G(A\otimes B))_m=(A\otimes B)_{ij}\nonumber\\
    &=A_{i//2^q,j//2^q}B_{i\%2^q,j\%2^q}
\end{align}
The last equality comes from the definition of tensor products. Here, the $\%$ symbol represents the remainder operator.
\begin{align}
    RHS&=(G(A)\otimes G(B))_{l(i,j)}\nonumber\\
    &=(G(A))_{l(i,j)//2^{2q}}(G(B))_{l(i,j)\%2^{2q}}\label{leq1}\\
    &=(G(A))_{l(i//2^{q},j//2^{q})}(G(B))_{l(i\%2^{q},j\%2^{q})}\label{leq2}\\
    &=A_{i//2^q,j//2^q}B_{i\%2^q,j\%2^q}\nonumber
\end{align}

To explain how we get from \eqref{leq1} to \eqref{leq2}, we note that $l(i,j)//2^{2q}$ is what we get when we interleave the bitstring representations of $i$ and $j$ and then get rid of the $2q$ least significant bits. We can get the same result by first getting rid of the $q$ least significant bits from both $i$ and $j$ and then interleaving their bits. This shows that $l(i,j)//2^{2q}=l(i//2^{q},j//2^{q})$. A similar explanation holds for $l(i,j)\%2^{2q}=l(i//2^{q},j\%2^{q})$.

Next we need to prepare an amplitude encoding of $M_{flat}$. While any state preparation approach would work, we use the Quantum Pixel representation (QPIXL) \cite{qpixl} to give us the following state (this is done based on the Flexible Representation of Quantum Images (FRQI) encoding \cite{frqi}, except we encode our matrix entries using amplitudes instead of angles. In FRQI, we would set each angle $\theta_i$ to be equal to an element $a_i$ up to some factor, whereas over here we set $\theta_i=2\arccos{a_i}$):

\begin{align}
    &\frac{1}{2^n}|M_{flat}\rangle_{dat}|0\rangle_{aux}+|\dots\rangle_{dat}|1\rangle_{aux}\nonumber\\
    &=\frac{1}{2^n}\sum_{\Vec{s}}c^M_{\Vec{s}}|P_{\Vec{s},flat}\rangle_{dat}|0\rangle_{aux}+|\dots\rangle_{dat}|1\rangle_{aux}\nonumber\\
    &=\frac{1}{2^n}\sum_{\Vec{s}}c^M_{\Vec{s}}(|P_{s_1,flat}\rangle\otimes\dots\otimes |P_{s_n,flat}\rangle)_{dat}
    |0\rangle_{aux}\nonumber\\
    &+|\dots\rangle_{dat}|1\rangle_{aux}\label{frqi}
\end{align}

where $|M_{flat}\rangle$, $|P_{\Vec{s},flat}\rangle$, $|P_{i,flat}\rangle$ represent flattened unnormalized statevectors of $M$, $P_{\Vec{s}}$, and $P_i$ respectively; and $M=\sum_{\Vec{s}}c^M_{\Vec{s}}P_{\Vec{s}}$.

\subsubsection{$\{|P_{i,flat}\rangle\}$ and Bell states}
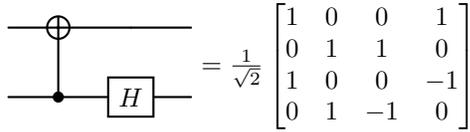
\begin {figure}[htbp]
\centering
\begin{quantikz}
\\
 & \targ{} & & \\
 & \ctrl{-1} & \gate{H} & \\
\end{quantikz}
$=\frac{1}{\sqrt{2}}
\begin{bmatrix}
    1 & 0 & 0 & 1 \\
    0 & 1 & 1 & 0 \\
    1 & 0 & 0 & -1 \\
    0 & 1 & -1 & 0
\end{bmatrix}$
\caption{Circuit B}
\label{fig:B}
\end{figure}

Closer inspection of $\{|P_{i,flat}\rangle\}$ reveals that these are the Bell states up to a scaling factor.
We can use circuit $B$ from Fig. \ref{fig:B} to go from the Bell basis to the computational basis.

\begin{align*}
    B|P_{0,flat}\rangle=\sqrt{2}|00\rangle\equiv\sqrt{2}|0\rangle\\
    B|P_{1,flat}\rangle=\sqrt{2}|01\rangle\equiv\sqrt{2}|1\rangle\\
    B|P_{2,flat}\rangle=\sqrt{2}|11\rangle\equiv\sqrt{2}|3\rangle\\
    B|P_{3,flat}\rangle=\sqrt{2}|10\rangle\equiv\sqrt{2}|2\rangle\\
\end{align*}

Let $g$ be the map that takes $0,1,2,3$ to $0,1,3,2$ respectively.

\begin{equation}
    B|P_{i,flat}\rangle=\sqrt{2}|g(i)\rangle
\end{equation}

Taking the amplitude encoded state that we had prepared in \eqref{frqi}, we apply $B$ to every consecutive pair of qubits in $dat$ to get the following state:

\begin{equation}
    \frac{1}{\sqrt{2^n}}\sum_{\Vec{s}}c^M_{\Vec{s}}(|g(s_1)\rangle\otimes\dots\otimes |g(s_n)\rangle)_{dat}
    |0\rangle_{aux}+|\dots\rangle_{dat}|1\rangle_{aux}
\end{equation}

\subsubsection{Applying $P_{\Vec{s}}$ on our input state}
\begin {figure}[htbp]
\centering
\begin{quantikz}
\\
 \lstick[2]{$|g(i)\rangle$} & \ctrl{2} & & \\
 & & \ctrl{1} & \\
 \lstick{$|\phi\rangle$} & \gate{X} & \gate{Z}
\end{quantikz}
\caption{Circuit K}
\label{fig:K}
\end{figure}
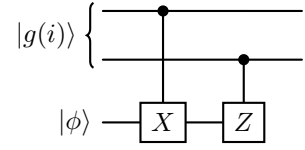

The circuit $K$ in Fig. \ref{fig:K} takes a state $|g(i)\rangle|\phi\rangle$ and returns the state $|g(i)\rangle(P_i|\phi\rangle)$.

Now, we prepare $|\psi\rangle$ on the $isr$ qubits and apply $K$ on each pair of qubits in $dat$ and the corresponding qubit in $isr$ to give us the following state

\begin{align}
    &\frac{1}{\sqrt{2^n}}\sum_{\Vec{s}}c^M_{\Vec{s}}(|g(s_1)\rangle\otimes\dots\otimes |g(s_n)\rangle)_{dat}\nonumber\\
    &(P_{s_1}\otimes\dots\otimes P_{s_n})|\psi\rangle_{isr}|0\rangle_{aux}+|\dots\rangle_{dat,isr}|1\rangle_{aux}\nonumber\\
    &=\frac{1}{\sqrt{2^n}}\sum_{\Vec{s}}c^M_{\Vec{s}}|g(s_1)\dots g(s_n)\rangle_{dat}P_{\Vec{s}}|\psi\rangle_{isr}|0\rangle_{aux}\nonumber\\
    &+|\dots\rangle_{dat,isr}|1\rangle_{aux}=:|\Phi\rangle\label{eq:final}
\end{align}

\subsubsection{Taking the inner product}
If we take the inner product of $|\Phi\rangle$ in \eqref{eq:final} and the following state

\begin{equation}
Q|0\rangle=|+\rangle^{\otimes 2n}_{dat}|0\rangle_{aux}|\psi\rangle_{isr},
\end{equation}

we get

\begin{align}
\langle0|Q^\dagger|\Phi\rangle&=\frac{1}{(2^n)^{3/2}}\sum_{\Vec{s}}c^M_{\Vec{s}}\langle\psi|_{isr}P_{\Vec{s}}|\psi\rangle_{isr}\nonumber\\
&=\frac{1}{(2^n)^{3/2}}\langle\psi|M|\psi\rangle.\label{innprod}
\end{align}

which is the quantity that we wanted. We can get $|\langle\psi|M|\psi\rangle|^2$ by preparing the state $Q^\dagger|\Phi\rangle$. The probability of measuring $|0\rangle_{dat}|0\rangle_{aux}|0\rangle_{isr}$ in this state is given by

\begin{equation}
    \text{Pr}(|0\rangle)=\frac{1}{2^{3n}}|\langle\psi|M|\psi\rangle|^2
\end{equation}

If we want to get the complex number $\langle\psi|M|\psi\rangle$ we can use Hadamard tests comparing $|\Phi\rangle$ and $Q|0\rangle$ to get its real and imaginary parts.
While we used QPIXL to prepare the initial statevector that encodes our matrix, we reiterate that any amplitude encoding would have worked. In general, we could have had an initial statevector like so:
\begin{equation}\label{geninitstate}
    \frac{1}{\eta}|M_{flat}\rangle_{dat}|0\rangle_{aux}+|\dots\rangle_{dat}|0^\perp\rangle_{aux}
\end{equation}
Here, $\eta$ is a positive real number and we allow $aux$ to represent a register of auxiliary qubits instead of just one qubit. Now, if we had initially been given such a statevector, \eqref{innprod} would have returned
\begin{equation}
    \frac{1}{\eta\sqrt{2^n}}\langle\psi|M|\psi\rangle.
\end{equation}
So in general, our method describes how to use a given prepared state \eqref{geninitstate} to construct a block encoding of $M$ with a subnormalization factor of $\alpha=\eta\sqrt{2^n}$, where it's worth noting that $\eta\geq||A||_F$.

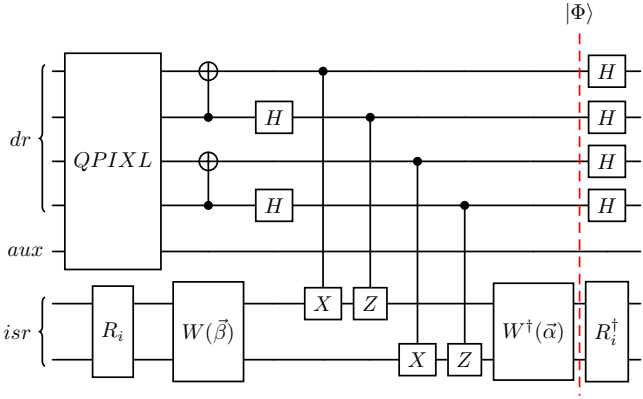
\begin {figure}[tb]
\centering
\resizebox{1.0\columnwidth}{!}{
\begin{quantikz}[row sep=0.2cm,column sep=0.2cm]
\\
 \lstick[4]{$dr$} & \gate[5]{QPIXL} & \targ{} & & \ctrl{5} & & & & \slice{$|\Phi\rangle$} & \gate{H} & \\
 & & \ctrl{-1} & \gate{H} & & \ctrl{4} & & & & \gate{H} & \\
 & & \targ{} & & & & \ctrl{4} & & & \gate{H} & \\
 & & \ctrl{-1} & \gate{H} & & & & \ctrl{3} & & \gate{H} & \\
 \lstick{$aux$} & & & & & & & & & & \\ 
 \lstick[2]{$isr$} & \gate[2]{R_i} & \gate[2]{W(\Vec{\beta})} & & \gate{X} & \gate{Z} & & & \gate[2]{W^\dagger(\Vec{\alpha})} & \gate[2]{R_i^\dagger} & \\
 & & & & & & \gate{X} & \gate{Z} & & &
\end{quantikz}
}
\caption{Full circuit for a $4\times 4$ matrix ($n=2$). The probability of measuring all the qubits to be $|0\rangle$ comes out to be $\frac{1}{2^{3n}}|\langle i|W^\dagger(\Vec{\alpha}) A W(\Vec{\beta})|i\rangle|^2$}
\label{fig:circfull}
\vspace{-5mm}
\end{figure}

\section{Implementing our variational quantum SVD algorithm}\label{sec:imp}

To compute our objective function, we need to evaluate $|\langle i|W^\dagger(\Vec{\alpha}) A W(\Vec{\beta})|i\rangle|^2$ for all $0\leq i\leq T$. To do so, we use our novel approach to evaluating expectation values from the last section, setting $|\psi\rangle=|i\rangle$ and $M=A$. The main change is that, when preparing our state $|\Phi\rangle$ from \eqref{eq:final}, we apply $W(\Vec{\beta})$ to $isr$ before we apply the $K$ operators and then we apply $W^\dagger(\Vec{\alpha})$ after. Let's define an operator $R_i$ such that $R_i|0\rangle=|i\rangle$. The final quantum circuit we use is depicted in Fig. \ref{fig:circfull}. As we can see, $7$ qubits are needed for $n=2$. In general, the number of qubits required for a $2^n\times2^n$ matrix is $3n+1$.

We use $T$ such circuits to evaluate the objective function when optimizing for $\Vec{\alpha}$ and $\Vec{\beta}$. At the end of the optimization we get our optimal parameters $\Vec{\alpha}*$ and $\Vec{\beta}*$. We set $\Tilde{U}=W(\Vec{\alpha}*)$ and $\Tilde{V}=W(\Vec{\beta}*)$. We then use Hadamard tests comparing $|\Phi\rangle$ and $|+\rangle^{\otimes 2n}_{dat}|0\rangle_{aux}|\psi\rangle_{isr}$ to get the real and imaginary parts of $\frac{1}{(2^n)^{3/2}}\langle i|\Tilde{U}^\dagger A\Tilde{V}|i\rangle$. We set $\Tilde{\sigma_i}=\langle i|\Tilde{U}^\dagger A\Tilde{V}|i\rangle$. The algorithm returns the two circuits $\Tilde{U}$ and $\Tilde{V}$ as well as a set of values $\{\Tilde{\sigma_i}\}$

\begin {figure}[tb]
\centering
$\dots$\begin{quantikz}[row sep=0.2cm,column sep=0.2cm]
 \gate[4]{L_{i-1}} & [0.3cm] \ctrl{1}\gategroup[4,steps=5]{$L_i$} & & & \gate{RZ(\gamma^i_{0})} & \gate{RY(\gamma^i_{4})} & [0.3cm] \gate[4]{L_{i+1}} \\
 & \targ{} & \ctrl{1} & & \gate{RZ(\gamma^i_1)} & \gate{RY(\gamma^i_{5})} & \\
 & & \targ{} & \ctrl{1} & \gate{RZ(\gamma^i_{2})} & \gate{RY(\gamma^i_{6})} & \\
 & & & \targ{} & \gate{RZ(\gamma^i_{3})} & \gate{RY(\gamma^i_{7})} &
\end{quantikz}$\dots$
\caption{Ansatz for $W(\Vec{\gamma})$ for a $16\times 16$ matrix ($n=4$ qubits). If the ansatz has $l$ layers, then $\Vec{\gamma}$ will consist of $2nl=8l$ individual parameters}
\label{fig:ansatz}
\vspace{-5mm}
\end{figure}
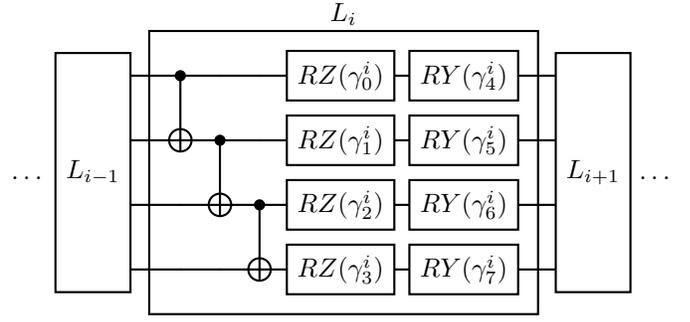

The parameterized circuit $W$ that we use is depicted in Fig. \ref{fig:ansatz}.

\begin{table}[b]
\centering
\caption{Circuit depth and gate counts for original objective function}
\resizebox{1.0\columnwidth}{!}{
\begin{tabular}{ccccccccc}
\toprule
\begin{tabular}{@{}c@{}}\textbf{Input}\\ \textbf{size}\end{tabular} & \begin{tabular}{@{}c@{}}\textbf{\# ansatz}\\ \textbf{layers}\end{tabular} & \textbf{Depth} & \textbf{H} & \textbf{CZ} & \textbf{CRY} & \textbf{CRZ} & \textbf{CH} & \textbf{CCX}\\
\midrule
\multirow{3}{*}{$2\times 2$} & 2 & 22      & 4     & 1    & 8     & 4    & 1   & 6     \\
 & 3 & 26       & 4     & 1      & 10    & 6      & 1       & 6        \\ 
                     & 4 & 30                     & 4          & 1           & 12           & 8            & 1           & 6                \\ 
\midrule
\multirow{3}{*}{$4\times 4$} & 2 & 62                     & 6          & 2           & 24           & 8            & 2           & 24               \\ 
                     & 3 & 72                     & 6          & 2           & 28           & 12           & 2           & 26               \\
                     & 4 & 82                     & 6          & 2           & 32           & 16           & 2           & 28               \\
\midrule
\multirow{4}{*}{$8\times 8$} & 2 & 174                    & 8          & 3           & 76           & 12           & 3           & 78               \\ 
                     & 3 & 190                    & 8          & 3           & 82           & 18           & 3           & 82               \\
                     & 4 & 206                    & 8          & 3           & 88           & 24           & 3           & 86               \\
                     & 5 & 222                    & 8          & 3           & 94           & 30           & 3           & 90               \\
                    
\bottomrule
\end{tabular}
}
\label{table:orig}
\end{table}

\setlength{\tabcolsep}{8.4pt}
\begin{table}[b]
\centering
\caption{Circuit depth and gate counts for new modified objective function}
\resizebox{1.0\columnwidth}{!}{
\begin{tabular}{cccccccc}

\toprule
\begin{tabular}{@{}c@{}}\textbf{Input}\\ \textbf{size}\end{tabular} & \begin{tabular}{@{}c@{}}\textbf{\# ansatz}\\ \textbf{layers}\end{tabular} & \textbf{Depth} & \textbf{H} & \textbf{RZ} & \textbf{RY} & \textbf{CX} & \textbf{CZ}\\

\midrule
\multirow{3}{*}{$2\times2$} & 2 & 15                     & 5          & 4           & 8           & 6             & 1           \\
                     & 3 & 17                     & 5          & 6           & 10          & 6             & 1           \\
                     & 4 & 19                     & 5          & 8           & 12          & 6             & 1           \\
\midrule
\multirow{3}{*}{$4\times4$} & 2 & 41                     & 10         & 8           & 24          & 24            & 2           \\
                     & 3 & 44                     & 10         & 12          & 28          & 26            & 2           \\ 
                     & 4 & 47                     & 10         & 16          & 32          & 28            & 2           \\
\midrule
\multirow{4}{*}{$8\times8$} & 2 & 138                    & 15         & 12          & 76          & 78            & 3           \\
                     & 3 & 142                    & 15         & 18          & 82          & 82            & 3           \\
                     & 4 & 146                    & 15         & 24          & 88          & 86            & 3           \\
                     & 5 & 150                    & 15         & 30          & 94          & 90            & 3           \\
                     
\bottomrule
\end{tabular}
}
\label{table:new}
\end{table}

Ideally we want an ansatz that should be able to assume any general unitary operator. The ansatz we're using may not be able to take the form of any general unitary; but as the number of layers increases, the closer it gets to that ideal.\\
However, the tradeoff is that the number of layers, while increasing the depth of the circuits, also increases the number of parameters and therefore the expected number of circuit runs it takes to get to the optimum.\\

For classical optimization, we use $\texttt{scipy.optimize.minimize}$ \cite{scipy} with the $\texttt{BFGS}$ (Broyden–Fletcher–Goldfarb–Shanno) \cite{BFGS} method. As mentioned earlier, we use QPIXL to get the encoding for $A_{flat}$.\\

\begin{figure}
    \centering
    \hspace*{-0.6cm}
\begin{tikzpicture}[scale=0.45]

\definecolor{darkgray176}{RGB}{176,176,176}
\definecolor{darkorange25512714}{RGB}{255,127,14}
\definecolor{forestgreen4416044}{RGB}{44,160,44}
\definecolor{lightgray204}{RGB}{204,204,204}
\definecolor{steelblue31119180}{RGB}{31,119,180}

\begin{groupplot}[group style={group size=2 by 3, horizontal sep=2cm, vertical sep=3cm}]
\nextgroupplot[
legend cell align={left},
legend style={fill opacity=0.8, draw opacity=1, text opacity=1, draw=lightgray204, nodes={scale=1.3, transform shape}},
tick align=outside,
tick pos=left,
x grid style={darkgray176},
xlabel={T},
xmin=0.95, xmax=2.05,
xtick distance={1},
xtick style={color=black},
y grid style={darkgray176},
ylabel={MSE},
ymin=-0.00165697405233966, ymax=0.0347964556759318,
ytick style={color=black}
]
\addplot [draw=steelblue31119180, fill=steelblue31119180, forget plot, mark=*, only marks]
table{%
x  y
1 0.033139481597374
2 1.46068860255143e-10
};
\path [fill=steelblue31119180, fill opacity=0.3]
(axis cs:1,0.0331394818544521)
--(axis cs:1,0.0331394813402958)
--(axis cs:2,-3.5425360116252e-10)
--(axis cs:2,6.46391321672806e-10)
--(axis cs:2,6.46391321672806e-10)
--(axis cs:1,0.0331394818544521)
--cycle;

\addplot [draw=darkorange25512714, fill=darkorange25512714, forget plot, mark=*, only marks]
table{%
x  y
1 0.0262552094829522
2 4.26399960675298e-11
};
\path [fill=darkorange25512714, fill opacity=0.3]
(axis cs:1,0.0262552097152587)
--(axis cs:1,0.0262552092506457)
--(axis cs:2,-1.08703694309686e-11)
--(axis cs:2,9.61503615660281e-11)
--(axis cs:2,9.61503615660281e-11)
--(axis cs:1,0.0262552097152587)
--cycle;

\addplot [draw=forestgreen4416044, fill=forestgreen4416044, forget plot, mark=*, only marks]
table{%
x  y
1 0.0198881616333893
2 2.62181365726537e-11
};
\path [fill=forestgreen4416044, fill opacity=0.3]
(axis cs:1,0.0198881617157659)
--(axis cs:1,0.0198881615510128)
--(axis cs:2,-1.06848907837445e-11)
--(axis cs:2,6.31211639290519e-11)
--(axis cs:2,6.31211639290519e-11)
--(axis cs:1,0.0198881617157659)
--cycle;

\addplot [semithick, steelblue31119180]
table {%
1 0.033139481597374
2 1.46068860255143e-10
};
\addlegendentry{2-layer ansatz}
\addplot [semithick, darkorange25512714]
table {%
1 0.0262552094829522
2 4.26399960675298e-11
};
\addlegendentry{3-layer ansatz}
\addplot [semithick, forestgreen4416044]
table {%
1 0.0198881616333893
2 2.62181365726537e-11
};
\addlegendentry{4-layer ansatz}

\nextgroupplot[
legend cell align={left},
legend style={fill opacity=0.8, draw opacity=1, text opacity=1, draw=lightgray204, nodes={scale=1.3, transform shape}},
tick align=outside,
tick pos=left,
x grid style={darkgray176},
xlabel={T},
xmin=0.95, xmax=2.05,
xtick distance={1},
xtick style={color=black},
y grid style={darkgray176},
ylabel={MSE},
ymin=-0.00132229431657736, ymax=0.0277681874178145,
ytick style={color=black}
]
\addplot [draw=steelblue31119180, fill=steelblue31119180, forget plot, mark=*, only marks]
table{%
x  y
1 0.0264458927935239
2 9.27363583938963e-10
};
\path [fill=steelblue31119180, fill opacity=0.3]
(axis cs:1,0.0264458932737901)
--(axis cs:1,0.0264458923132577)
--(axis cs:2,-8.31955939253856e-10)
--(axis cs:2,2.68668310713178e-09)
--(axis cs:2,2.68668310713178e-09)
--(axis cs:1,0.0264458932737901)
--cycle;

\addplot [draw=darkorange25512714, fill=darkorange25512714, forget plot, mark=*, only marks]
table{%
x  y
1 0.00943007391336767
2 3.0771317375754e-10
};
\path [fill=darkorange25512714, fill opacity=0.3]
(axis cs:1,0.00943007452098942)
--(axis cs:1,0.00943007330574593)
--(axis cs:2,-1.37698070186212e-10)
--(axis cs:2,7.53124417701293e-10)
--(axis cs:2,7.53124417701293e-10)
--(axis cs:1,0.00943007452098942)
--cycle;

\addplot [draw=forestgreen4416044, fill=forestgreen4416044, forget plot, mark=*, only marks]
table{%
x  y
1 0.0248315802572263
2 3.13672327882483e-10
};
\path [fill=forestgreen4416044, fill opacity=0.3]
(axis cs:1,0.0248315808646229)
--(axis cs:1,0.0248315796498297)
--(axis cs:2,-1.0675261791902e-10)
--(axis cs:2,7.34097273683987e-10)
--(axis cs:2,7.34097273683987e-10)
--(axis cs:1,0.0248315808646229)
--cycle;

\addplot [semithick, steelblue31119180]
table {%
1 0.0264458927935239
2 9.27363583938963e-10
};
\addlegendentry{2-layer ansatz}
\addplot [semithick, darkorange25512714]
table {%
1 0.00943007391336767
2 3.0771317375754e-10
};
\addlegendentry{3-layer ansatz}
\addplot [semithick, forestgreen4416044]
table {%
1 0.0248315802572263
2 3.13672327882483e-10
};
\addlegendentry{4-layer ansatz}

\nextgroupplot[
legend cell align={left},
legend style={fill opacity=0.8, draw opacity=1, text opacity=1, draw=lightgray204, nodes={scale=1.3, transform shape}},
tick align=outside,
tick pos=left,
x grid style={darkgray176},
xlabel={T},
xtick distance={1},
xmin=0.85, xmax=4.15,
xtick style={color=black},
y grid style={darkgray176},
ylabel={MSE},
ymin=-0.00267792550597809, ymax=0.0562558479645156,
ytick style={color=black}
]
\addplot [draw=steelblue31119180, fill=steelblue31119180, forget plot, mark=*, only marks]
table{%
x  y
1 0.0456788015128934
2 0.0363086168934577
3 0.0270914721584479
4 0.0266358543936898
};
\path [fill=steelblue31119180, fill opacity=0.3]
(axis cs:1,0.0458766057345585)
--(axis cs:1,0.0454809972912283)
--(axis cs:2,0.032341247949625)
--(axis cs:3,0.019966210061384)
--(axis cs:4,0.019422477192854)
--(axis cs:4,0.0338492315945257)
--(axis cs:4,0.0338492315945257)
--(axis cs:3,0.0342167342555118)
--(axis cs:2,0.0402759858372905)
--(axis cs:1,0.0458766057345585)
--cycle;

\addplot [draw=darkorange25512714, fill=darkorange25512714, forget plot, mark=*, only marks]
table{%
x  y
1 0.0535770400794932
2 0.0160140567923559
3 0.00307551565162399
4 0.00178266653624
};
\path [fill=darkorange25512714, fill opacity=0.3]
(axis cs:1,0.0535776872307124)
--(axis cs:1,0.053576392928274)
--(axis cs:2,0.0155347843085779)
--(axis cs:3,0.000820162422317637)
--(axis cs:4,-0.00072844001940448)
--(axis cs:4,0.00429377309188449)
--(axis cs:4,0.00429377309188449)
--(axis cs:3,0.00533086888093035)
--(axis cs:2,0.0164933292761339)
--(axis cs:1,0.0535776872307124)
--cycle;

\addplot [draw=forestgreen4416044, fill=forestgreen4416044, forget plot, mark=*, only marks]
table{%
x  y
1 0.0405551363191559
2 0.0126298257985957
3 0.0020523200818934
4 8.82379044350538e-07
};
\addplot [draw=forestgreen4416044, fill=forestgreen4416044, forget plot, mark=*, only marks]
table{%
x  y
1 0.0405551363191559
2 0.0126298257985957
3 0.0020523200818934
4 8.82379044350538e-07
};
\path [fill=forestgreen4416044, fill opacity=0.3]
(axis cs:1,0.0405551368206954)
--(axis cs:1,0.0405551358176163)
--(axis cs:2,0.0126298067288185)
--(axis cs:3,0.00204122684780773)
--(axis cs:4,-7.12043422900133e-06)
--(axis cs:4,8.88519231770241e-06)
--(axis cs:4,8.88519231770241e-06)
--(axis cs:3,0.00206341331597908)
--(axis cs:2,0.012629844868373)
--(axis cs:1,0.0405551368206954)
--cycle;

\addplot [semithick, steelblue31119180]
table {%
1 0.0456788015128934
2 0.0363086168934577
3 0.0270914721584479
4 0.0266358543936898
};
\addlegendentry{2-layer ansatz}
\addplot [semithick, darkorange25512714]
table {%
1 0.0535770400794932
2 0.0160140567923559
3 0.00307551565162399
4 0.00178266653624
};
\addlegendentry{3-layer ansatz}
\addplot [semithick, forestgreen4416044]
table {%
1 0.0405551363191559
2 0.0126298257985957
3 0.0020523200818934
4 8.82379044350538e-07
};
\addlegendentry{4-layer ansatz}

\nextgroupplot[
legend cell align={left},
legend style={fill opacity=0.8, draw opacity=1, text opacity=1, draw=lightgray204, nodes={scale=1.3, transform shape}},
tick align=outside,
tick pos=left,
x grid style={darkgray176},
xlabel={T},
xmin=0.85, xmax=4.15,
xtick distance={1},
xtick style={color=black},
y grid style={darkgray176},
ylabel={MSE},
ymin=-0.00264010455377724, ymax=0.0561237771006405,
ytick style={color=black}
]
\addplot [draw=steelblue31119180, fill=steelblue31119180, forget plot, mark=*, only marks]
table{%
x  y
1 0.0401306860403062
2 0.0204564029141119
3 0.0145747165019856
4 0.0117767905578406
};
\path [fill=steelblue31119180, fill opacity=0.3]
(axis cs:1,0.0413034813099185)
--(axis cs:1,0.0389578907706939)
--(axis cs:2,0.0138305631235102)
--(axis cs:3,0.00780255038429333)
--(axis cs:4,0.00494040513328224)
--(axis cs:4,0.0186131759823991)
--(axis cs:4,0.0186131759823991)
--(axis cs:3,0.0213468826196779)
--(axis cs:2,0.0270822427047136)
--(axis cs:1,0.0413034813099185)
--cycle;

\addplot [draw=darkorange25512714, fill=darkorange25512714, forget plot, mark=*, only marks]
table{%
x  y
1 0.0534526915708943
2 0.0171599129336688
3 0.0030323111159876
4 0.000986275554263904
};
\path [fill=darkorange25512714, fill opacity=0.3]
(axis cs:1,0.0534538868781669)
--(axis cs:1,0.0534514962636216)
--(axis cs:2,0.0134777317118142)
--(axis cs:3,0.00103513470532919)
--(axis cs:4,-0.000335399735217057)
--(axis cs:4,0.00230795084374486)
--(axis cs:4,0.00230795084374486)
--(axis cs:3,0.005029487526646)
--(axis cs:2,0.0208420941555235)
--(axis cs:1,0.0534538868781669)
--cycle;

\addplot [draw=forestgreen4416044, fill=forestgreen4416044, forget plot, mark=*, only marks]
table{%
x  y
1 0.0424132268764088
2 0.0125440026832668
3 0.00111414559611261
4 3.09809759690259e-05
};
\path [fill=forestgreen4416044, fill opacity=0.3]
(axis cs:1,0.0424132401204949)
--(axis cs:1,0.0424132136323227)
--(axis cs:2,0.0125419856621531)
--(axis cs:3,0.000949307395195137)
--(axis cs:4,-9.78266516195078e-05)
--(axis cs:4,0.00015978860355756)
--(axis cs:4,0.00015978860355756)
--(axis cs:3,0.00127898379703008)
--(axis cs:2,0.0125460197043805)
--(axis cs:1,0.0424132401204949)
--cycle;

\addplot [semithick, steelblue31119180]
table {%
1 0.0401306860403062
2 0.0204564029141119
3 0.0145747165019856
4 0.0117767905578406
};
\addlegendentry{2-layer ansatz}
\addplot [semithick, darkorange25512714]
table {%
1 0.0534526915708943
2 0.0171599129336688
3 0.0030323111159876
4 0.000986275554263904
};
\addlegendentry{3-layer ansatz}
\addplot [semithick, forestgreen4416044]
table {%
1 0.0424132268764088
2 0.0125440026832668
3 0.00111414559611261
4 3.09809759690259e-05
};
\addlegendentry{4-layer ansatz}

\nextgroupplot[
legend cell align={left},
legend style={fill opacity=0.5, draw opacity=0.5, text opacity=1,
  at={(0.03,0.97)},
  anchor=north west, draw=lightgray204, nodes={scale=1.3, transform shape}},
tick align=outside,
tick pos=left,
x grid style={darkgray176},
xlabel={T},
xmin=0.65, xmax=8.35,
xtick distance={1},
xtick style={color=black},
y grid style={darkgray176},
ylabel={MSE},
ymin=0.0302606475654334, ymax=0.086276613518792,
ytick style={color=black}
]
\addplot [draw=steelblue31119180, fill=steelblue31119180, forget plot, mark=*, only marks]
table{%
x  y
1 0.0628240365024877
2 0.0583439936283316
3 0.0553240120750498
4 0.0583524057072647
5 0.059120539099129
6 0.0587012240962353
7 0.063508968604108
8 0.0634849431948027
};
\path [fill=steelblue31119180, fill opacity=0.3]
(axis cs:1,0.0647651826335022)
--(axis cs:1,0.0608828903714732)
--(axis cs:2,0.0544685661878063)
--(axis cs:3,0.0507305130707357)
--(axis cs:4,0.0537968152978416)
--(axis cs:5,0.0521631313729909)
--(axis cs:6,0.0508927938457443)
--(axis cs:7,0.0526693120469349)
--(axis cs:8,0.0488793641878276)
--(axis cs:8,0.0780905222017779)
--(axis cs:8,0.0780905222017779)
--(axis cs:7,0.0743486251612812)
--(axis cs:6,0.0665096543467263)
--(axis cs:5,0.0660779468252671)
--(axis cs:4,0.0629079961166877)
--(axis cs:3,0.059917511079364)
--(axis cs:2,0.0622194210688569)
--(axis cs:1,0.0647651826335022)
--cycle;
\path [fill=darkorange25512714, fill opacity=0.3]
(axis cs:1,0.0629210039869922)
--(axis cs:1,0.0616306513483299)
--(axis cs:2,0.0453676052500166)
--(axis cs:3,0.0427153215184374)
--(axis cs:4,0.0417271683492162)
--(axis cs:5,0.0399007927673713)
--(axis cs:6,0.0415213591609364)
--(axis cs:7,0.043043398858284)
--(axis cs:8,0.0423903581510904)
--(axis cs:8,0.071254794334051)
--(axis cs:8,0.071254794334051)
--(axis cs:7,0.0656975907278905)
--(axis cs:6,0.0567183540583423)
--(axis cs:5,0.0517185979429264)
--(axis cs:4,0.0514047200698322)
--(axis cs:3,0.0511316040634337)
--(axis cs:2,0.0525618866630841)
--(axis cs:1,0.0629210039869922)
--cycle;

\addplot [draw=darkorange25512714, fill=darkorange25512714, forget plot, mark=*, only marks]
table{%
x  y
1 0.062275827667661
2 0.0489647459565503
3 0.0469234627909355
4 0.0465659442095242
5 0.0458096953551489
6 0.0491198566096393
7 0.0543704947930872
8 0.0568225762425707
};
\path [fill=forestgreen4416044, fill opacity=0.3]
(axis cs:1,0.0646402892485596)
--(axis cs:1,0.0646387590659918)
--(axis cs:2,0.0385040709558809)
--(axis cs:3,0.0304155572637646)
--(axis cs:4,0.0285896478430414)
--(axis cs:5,0.027087020840948)
--(axis cs:6,0.027618864840832)
--(axis cs:7,0.0282104217842031)
--(axis cs:8,0.0289399933217185)
--(axis cs:8,0.0464010190459554)
--(axis cs:8,0.0464010190459554)
--(axis cs:7,0.0439667876781477)
--(axis cs:6,0.0394271875214942)
--(axis cs:5,0.0367084530129514)
--(axis cs:4,0.0363594190906279)
--(axis cs:3,0.0374788668826323)
--(axis cs:2,0.0413268113856372)
--(axis cs:1,0.0646402892485596)
--cycle;

\addplot [draw=forestgreen4416044, fill=forestgreen4416044, forget plot, mark=*, only marks]
table{%
x  y
1 0.0646395241572757
2 0.039915441170759
3 0.0339472120731985
4 0.0324745334668347
5 0.0318977369269497
6 0.0335230261811631
7 0.0360886047311754
8 0.0376705061838369
};
\addplot [semithick, steelblue31119180]
table {%
1 0.0628240365024877
2 0.0583439936283316
3 0.0553240120750498
4 0.0583524057072647
5 0.059120539099129
6 0.0587012240962353
7 0.063508968604108
8 0.0634849431948027
};
\addlegendentry{2-layer ansatz}
\addplot [semithick, darkorange25512714]
table {%
1 0.062275827667661
2 0.0489647459565503
3 0.0469234627909355
4 0.0465659442095242
5 0.0458096953551489
6 0.0491198566096393
7 0.0543704947930872
8 0.0568225762425707
};
\addlegendentry{3-layer ansatz}
\addplot [semithick, forestgreen4416044]
table {%
1 0.0646395241572757
2 0.039915441170759
3 0.0339472120731985
4 0.0324745334668347
5 0.0318977369269497
6 0.0335230261811631
7 0.0360886047311754
8 0.0376705061838369
};
\addlegendentry{4-layer ansatz}

\nextgroupplot[
legend cell align={left},
legend style={fill opacity=0.8, draw opacity=1, text opacity=1, draw=lightgray204, nodes={scale=1.3, transform shape}},
tick align=outside,
tick pos=left,
x grid style={darkgray176},
xlabel={T},
xmin=0.65, xmax=8.35,
xtick distance={1},
xtick style={color=black},
y grid style={darkgray176},
ylabel={MSE},
ymin=0.0178700997187847, ymax=0.0862579487814735,
ytick style={color=black}
]
\addplot [draw=steelblue31119180, fill=steelblue31119180, forget plot, mark=*, only marks]
table{%
x  y
1 0.0697857738240786
2 0.0679527783656023
3 0.0648084179102343
4 0.0634460696624174
5 0.0608218555242107
6 0.0570238970053954
7 0.0570341505838385
8 0.0548370029110403
};
\path [fill=steelblue31119180, fill opacity=0.3]
(axis cs:1,0.0710348260225162)
--(axis cs:1,0.0685367216256409)
--(axis cs:2,0.0661514998779505)
--(axis cs:3,0.0621196220051697)
--(axis cs:4,0.061092298450325)
--(axis cs:5,0.0524952575508183)
--(axis cs:6,0.0531559115432412)
--(axis cs:7,0.0454891812885976)
--(axis cs:8,0.0446689130716838)
--(axis cs:8,0.0650050927503969)
--(axis cs:8,0.0650050927503969)
--(axis cs:7,0.0685791198790795)
--(axis cs:6,0.0608918824675497)
--(axis cs:5,0.0691484534976031)
--(axis cs:4,0.0657998408745097)
--(axis cs:3,0.067497213815299)
--(axis cs:2,0.0697540568532541)
--(axis cs:1,0.0710348260225162)
--cycle;

\addplot [draw=darkorange25512714, fill=darkorange25512714, forget plot, mark=*, only marks]
table{%
x  y
1 0.0608250093079127
2 0.0527208031637763
3 0.0438495509613603
4 0.0404483830214799
5 0.0376166709213337
6 0.0356538870559485
7 0.0335370104406352
8 0.0324630905270775
};
\path [fill=darkorange25512714, fill opacity=0.3]
(axis cs:1,0.0620181062897076)
--(axis cs:1,0.0596319123261177)
--(axis cs:2,0.0497363403066033)
--(axis cs:3,0.0395507125833832)
--(axis cs:4,0.0360966797387788)
--(axis cs:5,0.0332138783503932)
--(axis cs:6,0.0305269344556773)
--(axis cs:7,0.0289549192553684)
--(axis cs:8,0.027827257001415)
--(axis cs:8,0.0370989240527401)
--(axis cs:8,0.0370989240527401)
--(axis cs:7,0.038119101625902)
--(axis cs:6,0.0407808396562198)
--(axis cs:5,0.0420194634922742)
--(axis cs:4,0.044800086304181)
--(axis cs:3,0.0481483893393375)
--(axis cs:2,0.0557052660209494)
--(axis cs:1,0.0620181062897076)
--cycle;

\addplot [draw=forestgreen4416044, fill=forestgreen4416044, forget plot, mark=*, only marks]
table{%
x  y
1 0.0594717388664473
2 0.0420155208093759
3 0.0318506423875343
4 0.0284122898069905
5 0.0256830235877478
6 0.0232210521849409
7 0.0213801705367183
8 0.0203422746761796
};
\path [fill=forestgreen4416044, fill opacity=0.3]
(axis cs:1,0.0594848672898723)
--(axis cs:1,0.0594586104430223)
--(axis cs:2,0.0385228080595874)
--(axis cs:3,0.0277176931154235)
--(axis cs:4,0.0256527650957839)
--(axis cs:5,0.0220635840997145)
--(axis cs:6,0.0200585811057919)
--(axis cs:7,0.0185933340062491)
--(axis cs:8,0.0175585275459191)
--(axis cs:8,0.0231260218064401)
--(axis cs:8,0.0231260218064401)
--(axis cs:7,0.0241670070671876)
--(axis cs:6,0.0263835232640899)
--(axis cs:5,0.0293024630757811)
--(axis cs:4,0.0311718145181971)
--(axis cs:3,0.0359835916596451)
--(axis cs:2,0.0455082335591645)
--(axis cs:1,0.0594848672898723)
--cycle;

\addplot [semithick, steelblue31119180]
table {%
1 0.0697857738240786
2 0.0679527783656023
3 0.0648084179102343
4 0.0634460696624174
5 0.0608218555242107
6 0.0570238970053954
7 0.0570341505838385
8 0.0548370029110403
};
\addlegendentry{2-layer ansatz}
\addplot [semithick, darkorange25512714]
table {%
1 0.0608250093079127
2 0.0527208031637763
3 0.0438495509613603
4 0.0404483830214799
5 0.0376166709213337
6 0.0356538870559485
7 0.0335370104406352
8 0.0324630905270775
};
\addlegendentry{3-layer ansatz}
\addplot [semithick, forestgreen4416044]
table {%
1 0.0594717388664473
2 0.0420155208093759
3 0.0318506423875343
4 0.0284122898069905
5 0.0256830235877478
6 0.0232210521849409
7 0.0213801705367183
8 0.0203422746761796
};
\addlegendentry{4-layer ansatz}
\end{groupplot}
\node[text width=6cm,align=center,anchor=north] at ($(group c1r1.south)-(0,1cm)$) {(a)};
\node[text width=6cm,align=center,anchor=north] at ($(group c2r1.south)-(0,1cm)$) {(d)};
\node[text width=6cm,align=center,anchor=north] at ($(group c1r2.south)-(0,1cm)$) {(b)};
\node[text width=6cm,align=center,anchor=north] at ($(group c2r2.south)-(0,1cm)$) {(e)};
\node[text width=6cm,align=center,anchor=north] at ($(group c1r3.south)-(0,1cm)$) {(c)};
\node[text width=6cm,align=center,anchor=north] at ($(group c2r3.south)-(0,1cm)$) {(f)};
\end{tikzpicture}
    \caption{The graphs depict the MSE of the matrix reconstructions when simulating the algorithm for randomly generated matrices of various sizes. A solid line represents the MSE averaged over all $100$ runs of the algorithm, and the shaded region around it represents $\pm 1$ standard deviation around the mean. (a), (b), and (c) are for $2\times2$, $4\times4$, and $8\times8$ matrices respectively using the original objective function. (d), (e), (f) are for $2\times2$, $4\times4$, and $8\times8$ matrices respectively using the modified objective function.}
    \label{fig:rand_mse}
\end{figure}
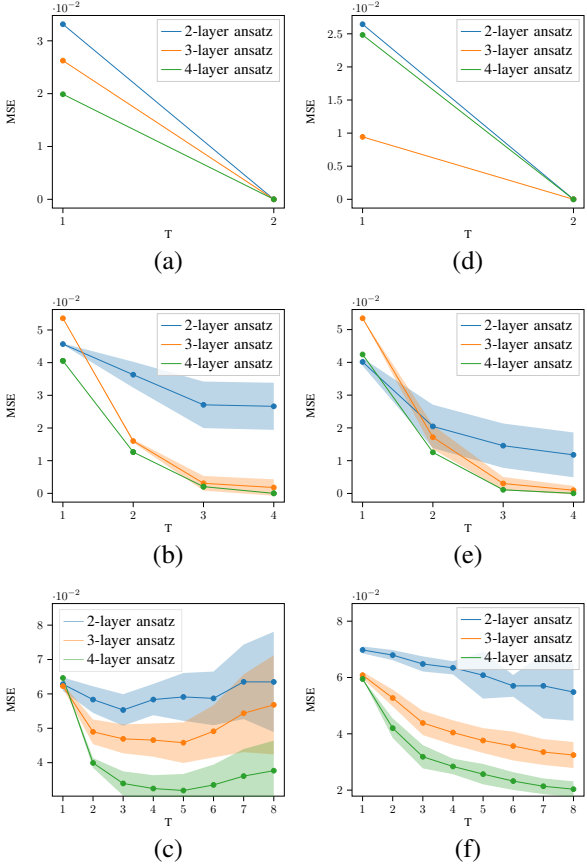

\begin{figure}
    \centering
    \hspace*{-0.6cm}
\begin{tikzpicture}[scale=0.45]

\definecolor{darkgray176}{RGB}{176,176,176}
\definecolor{darkorange25512714}{RGB}{255,127,14}
\definecolor{forestgreen4416044}{RGB}{44,160,44}
\definecolor{lightgray204}{RGB}{204,204,204}
\definecolor{steelblue31119180}{RGB}{31,119,180}

\begin{groupplot}[group style={group size=2 by 3, horizontal sep=2cm, vertical sep=3cm}]
\nextgroupplot[
legend cell align={left},
legend style={
  fill opacity=0.8,
  draw opacity=1,
  text opacity=1,
  at={(0.03,0.97)},
  anchor=north west,
  draw=lightgray204
},
tick align=outside,
tick pos=left,
x grid style={darkgray176},
xlabel={T},
xmin=0.95, xmax=2.05,
xtick distance={1},
xtick style={color=black},
y grid style={darkgray176},
ylabel={Circuit Runs},
ymin=173.796, ymax=433.044,
ytick style={color=black}
]
\addplot [draw=steelblue31119180, fill=steelblue31119180, forget plot, mark=*, only marks]
table{%
x  y
1 185.58
2 330.66
};
\path [fill=steelblue31119180, fill opacity=0.3]
(axis cs:1,219.035313479326)
--(axis cs:1,152.124686520674)
--(axis cs:2,269.187957574195)
--(axis cs:2,392.132042425805)
--(axis cs:2,392.132042425805)
--(axis cs:1,219.035313479326)
--cycle;

\addplot [draw=darkorange25512714, fill=darkorange25512714, forget plot, mark=*, only marks]
table{%
x  y
1 196.82
2 332.28
};
\path [fill=darkorange25512714, fill opacity=0.3]
(axis cs:1,231.591051177668)
--(axis cs:1,162.048948822332)
--(axis cs:2,289.33668853011)
--(axis cs:2,375.22331146989)
--(axis cs:2,375.22331146989)
--(axis cs:1,231.591051177668)
--cycle;

\addplot [draw=forestgreen4416044, fill=forestgreen4416044, forget plot, mark=*, only marks]
table{%
x  y
1 224.4
2 421.26
};
\path [fill=forestgreen4416044, fill opacity=0.3]
(axis cs:1,257.135485333198)
--(axis cs:1,191.664514666802)
--(axis cs:2,354.538519201085)
--(axis cs:2,487.981480798915)
--(axis cs:2,487.981480798915)
--(axis cs:1,257.135485333198)
--cycle;

\addplot [semithick, steelblue31119180]
table {%
1 185.58
2 330.66
};
\addlegendentry{2-layer ansatz}
\addplot [semithick, darkorange25512714]
table {%
1 196.82
2 332.28
};
\addlegendentry{3-layer ansatz}
\addplot [semithick, forestgreen4416044]
table {%
1 224.4
2 421.26
};
\addlegendentry{4-layer ansatz}

\nextgroupplot[
legend cell align={left},
legend style={
  fill opacity=0.8,
  draw opacity=1,
  text opacity=1,
  at={(0.03,0.97)},
  anchor=north west,
  draw=lightgray204
},
tick align=outside,
tick pos=left,
x grid style={darkgray176},
xlabel={T},
xmin=0.95, xmax=2.05,
xtick distance={1},
xtick style={color=black},
y grid style={darkgray176},
ylabel={Circuit Runs},
ymin=191.2815, ymax=606.3885,
ytick style={color=black}
]
\addplot [draw=steelblue31119180, fill=steelblue31119180, forget plot, mark=*, only marks]
table{%
x  y
1 210.15
2 448.56
};
\path [fill=steelblue31119180, fill opacity=0.3]
(axis cs:1,247.664997001199)
--(axis cs:1,172.635002998801)
--(axis cs:2,372.711835882468)
--(axis cs:2,524.408164117532)
--(axis cs:2,524.408164117532)
--(axis cs:1,247.664997001199)
--cycle;

\addplot [draw=darkorange25512714, fill=darkorange25512714, forget plot, mark=*, only marks]
table{%
x  y
1 240.5
2 485.68
};
\path [fill=darkorange25512714, fill opacity=0.3]
(axis cs:1,278.588003360638)
--(axis cs:1,202.411996639362)
--(axis cs:2,394.895684173972)
--(axis cs:2,576.464315826028)
--(axis cs:2,576.464315826028)
--(axis cs:1,278.588003360638)
--cycle;

\addplot [draw=forestgreen4416044, fill=forestgreen4416044, forget plot, mark=*, only marks]
table{%
x  y
1 259.93
2 587.52
};
\path [fill=forestgreen4416044, fill opacity=0.3]
(axis cs:1,299.975736352326)
--(axis cs:1,219.884263647674)
--(axis cs:2,482.780131754904)
--(axis cs:2,692.259868245096)
--(axis cs:2,692.259868245096)
--(axis cs:1,299.975736352326)
--cycle;

\addplot [semithick, steelblue31119180]
table {%
1 210.15
2 448.56
};
\addlegendentry{2-layer ansatz}
\addplot [semithick, darkorange25512714]
table {%
1 240.5
2 485.68
};
\addlegendentry{3-layer ansatz}
\addplot [semithick, forestgreen4416044]
table {%
1 259.93
2 587.52
};
\addlegendentry{4-layer ansatz}

\nextgroupplot[
legend cell align={left},
legend style={
  fill opacity=0.8,
  draw opacity=1,
  text opacity=1,
  at={(0.03,0.97)},
  anchor=north west,
  draw=lightgray204
},
tick align=outside,
tick pos=left,
x grid style={darkgray176},
xlabel={T},
xmin=0.85, xmax=4.15,
xtick distance={1},
xtick style={color=black},
y grid style={darkgray176},
ylabel={Circuit Runs},
ymin=69.0794999999999, ymax=17933.7505,
ytick style={color=black}
]
\addplot [draw=steelblue31119180, fill=steelblue31119180, forget plot, mark=*, only marks]
table{%
x  y
1 881.11
2 2343.62
3 2785.62
4 3692.4
};
\path [fill=steelblue31119180, fill opacity=0.3]
(axis cs:1,1027.46509215603)
--(axis cs:1,734.75490784397)
--(axis cs:2,1803.27131627809)
--(axis cs:3,2226.24973085799)
--(axis cs:4,2846.48983928552)
--(axis cs:4,4538.31016071448)
--(axis cs:4,4538.31016071448)
--(axis cs:3,3344.99026914201)
--(axis cs:2,2883.96868372191)
--(axis cs:1,1027.46509215603)
--cycle;

\addplot [draw=darkorange25512714, fill=darkorange25512714, forget plot, mark=*, only marks]
table{%
x  y
1 1311
2 6108.5
3 7455
4 9629
};
\path [fill=darkorange25512714, fill opacity=0.3]
(axis cs:1,1591.56639142991)
--(axis cs:1,1030.43360857009)
--(axis cs:2,4442.74716719436)
--(axis cs:3,5706.45739256946)
--(axis cs:4,7553.75302674597)
--(axis cs:4,11704.246973254)
--(axis cs:4,11704.246973254)
--(axis cs:3,9203.54260743054)
--(axis cs:2,7774.25283280564)
--(axis cs:1,1591.56639142991)
--cycle;

\addplot [draw=forestgreen4416044, fill=forestgreen4416044, forget plot, mark=*, only marks]
table{%
x  y
1 1069.86
2 5654.22
3 13213.53
4 17121.72
};
\path [fill=forestgreen4416044, fill opacity=0.3]
(axis cs:1,1207.11778666436)
--(axis cs:1,932.602213335636)
--(axis cs:2,4208.69795035842)
--(axis cs:3,9243.99416882276)
--(axis cs:4,12999.0766091645)
--(axis cs:4,21244.3633908355)
--(axis cs:4,21244.3633908355)
--(axis cs:3,17183.0658311772)
--(axis cs:2,7099.74204964158)
--(axis cs:1,1207.11778666436)
--cycle;

\addplot [semithick, steelblue31119180]
table {%
1 881.11
2 2343.62
3 2785.62
4 3692.4
};
\addlegendentry{2-layer ansatz}
\addplot [semithick, darkorange25512714]
table {%
1 1311
2 6108.5
3 7455
4 9629
};
\addlegendentry{3-layer ansatz}
\addplot [semithick, forestgreen4416044]
table {%
1 1069.86
2 5654.22
3 13213.53
4 17121.72
};
\addlegendentry{4-layer ansatz}


\nextgroupplot[
legend cell align={left},
legend style={
  fill opacity=0.8,
  draw opacity=1,
  text opacity=1,
  at={(0.03,0.97)},
  anchor=north west,
  draw=lightgray204
},
tick align=outside,
tick pos=left,
x grid style={darkgray176},
xlabel={T},
xmin=0.85, xmax=4.15,
xtick distance={1},
xtick style={color=black},
y grid style={darkgray176},
ylabel={Circuit Runs},
ymin=-116.565, ymax=28829.825,
ytick style={color=black}
]
\addplot [draw=steelblue31119180, fill=steelblue31119180, forget plot, mark=*, only marks]
table{%
x  y
1 1199.18
2 3613.52
3 5088.27
4 6656.52
};
\path [fill=steelblue31119180, fill opacity=0.3]
(axis cs:1,1391.16562967056)
--(axis cs:1,1007.19437032944)
--(axis cs:2,2996.37176497052)
--(axis cs:3,4194.87430659198)
--(axis cs:4,5467.73879557254)
--(axis cs:4,7845.30120442746)
--(axis cs:4,7845.30120442746)
--(axis cs:3,5981.66569340802)
--(axis cs:2,4230.66823502948)
--(axis cs:1,1391.16562967056)
--cycle;

\addplot [draw=darkorange25512714, fill=darkorange25512714, forget plot, mark=*, only marks]
table{%
x  y
1 1747.5
2 8614
3 13617.75
4 17884
};
\path [fill=darkorange25512714, fill opacity=0.3]
(axis cs:1,2055.46915429958)
--(axis cs:1,1439.53084570042)
--(axis cs:2,6819.09331718888)
--(axis cs:3,10712.6687834073)
--(axis cs:4,14407.1105855952)
--(axis cs:4,21360.8894144048)
--(axis cs:4,21360.8894144048)
--(axis cs:3,16522.8312165927)
--(axis cs:2,10408.9066828111)
--(axis cs:1,2055.46915429958)
--cycle;

\addplot [draw=forestgreen4416044, fill=forestgreen4416044, forget plot, mark=*, only marks]
table{%
x  y
1 1702.47
2 9614.22
3 21131.55
4 27514.08
};
\path [fill=forestgreen4416044, fill opacity=0.3]
(axis cs:1,1976.92407812601)
--(axis cs:1,1428.01592187399)
--(axis cs:2,7907.32550413917)
--(axis cs:3,17222.7784959338)
--(axis cs:4,22862.7588560668)
--(axis cs:4,32165.4011439332)
--(axis cs:4,32165.4011439332)
--(axis cs:3,25040.3215040662)
--(axis cs:2,11321.1144958608)
--(axis cs:1,1976.92407812601)
--cycle;

\addplot [semithick, steelblue31119180]
table {%
1 1199.18
2 3613.52
3 5088.27
4 6656.52
};
\addlegendentry{2-layer ansatz}
\addplot [semithick, darkorange25512714]
table {%
1 1747.5
2 8614
3 13617.75
4 17884
};
\addlegendentry{3-layer ansatz}
\addplot [semithick, forestgreen4416044]
table {%
1 1702.47
2 9614.22
3 21131.55
4 27514.08
};
\addlegendentry{4-layer ansatz}

\nextgroupplot[
legend cell align={left},
legend style={
  fill opacity=0.8,
  draw opacity=1,
  text opacity=1,
  at={(0.03,0.97)},
  anchor=north west,
  draw=lightgray204
},
tick align=outside,
tick pos=left,
x grid style={darkgray176},
xlabel={T},
xmin=0.65, xmax=8.35,
xtick distance={1},
xtick style={color=black},
y grid style={darkgray176},
ylabel={Circuit Runs},
ymin=981.5295, ymax=57746.3805,
ytick style={color=black}
]
\addplot [draw=steelblue31119180, fill=steelblue31119180, forget plot, mark=*, only marks]
table{%
x  y
1 3561.75
2 6024.5
3 7740.75
4 8237
5 10110
6 10612.5
7 12227.25
8 13606
};
\path [fill=steelblue31119180, fill opacity=0.3]
(axis cs:1,4766.01922861958)
--(axis cs:1,2357.48077138042)
--(axis cs:2,4791.64863020719)
--(axis cs:3,6200.45558820724)
--(axis cs:4,6511.14658791658)
--(axis cs:5,7704.46601562148)
--(axis cs:6,8190.73153460121)
--(axis cs:7,9464.19451286262)
--(axis cs:8,10707.1657515477)
--(axis cs:8,16504.8342484523)
--(axis cs:8,16504.8342484523)
--(axis cs:7,14990.3054871374)
--(axis cs:6,13034.2684653988)
--(axis cs:5,12515.5339843785)
--(axis cs:4,9962.85341208342)
--(axis cs:3,9281.04441179276)
--(axis cs:2,7257.35136979281)
--(axis cs:1,4766.01922861958)
--cycle;

\addplot [draw=darkorange25512714, fill=darkorange25512714, forget plot, mark=*, only marks]
table{%
x  y
1 7631.25
2 14751.9
3 17491.38
4 20295.24
5 21654.25
6 23625.24
7 26244.47
8 29674
};
\path [fill=darkorange25512714, fill opacity=0.3]
(axis cs:1,9444.73735286464)
--(axis cs:1,5817.76264713536)
--(axis cs:2,11412.158581866)
--(axis cs:3,13712.0329349635)
--(axis cs:4,15310.0054867599)
--(axis cs:5,16044.987154046)
--(axis cs:6,17827.0691767144)
--(axis cs:7,19506.2028327678)
--(axis cs:8,22534.8076692107)
--(axis cs:8,36813.1923307892)
--(axis cs:8,36813.1923307892)
--(axis cs:7,32982.7371672322)
--(axis cs:6,29423.4108232856)
--(axis cs:5,27263.512845954)
--(axis cs:4,25280.4745132401)
--(axis cs:3,21270.7270650365)
--(axis cs:2,18091.641418134)
--(axis cs:1,9444.73735286464)
--cycle;

\addplot [draw=forestgreen4416044, fill=forestgreen4416044, forget plot, mark=*, only marks]
table{%
x  y
1 6942.32
2 29938.02
3 33771.78
4 37392.88
5 40990.95
6 43853.04
7 49330.26
8 55166.16
};
\path [fill=forestgreen4416044, fill opacity=0.3]
(axis cs:1,8283.83762940336)
--(axis cs:1,5600.80237059665)
--(axis cs:2,23160.9044162136)
--(axis cs:3,25481.1676616983)
--(axis cs:4,28309.1945012611)
--(axis cs:5,30328.7340893649)
--(axis cs:6,33778.0413407445)
--(axis cs:7,37757.0110038019)
--(axis cs:8,40898.905468311)
--(axis cs:8,69433.414531689)
--(axis cs:8,69433.414531689)
--(axis cs:7,60903.5089961981)
--(axis cs:6,53928.0386592555)
--(axis cs:5,51653.1659106351)
--(axis cs:4,46476.5654987389)
--(axis cs:3,42062.3923383017)
--(axis cs:2,36715.1355837864)
--(axis cs:1,8283.83762940336)
--cycle;

\addplot [semithick, steelblue31119180]
table {%
1 3561.75
2 6024.5
3 7740.75
4 8237
5 10110
6 10612.5
7 12227.25
8 13606
};
\addlegendentry{2-layer ansatz}
\addplot [semithick, darkorange25512714]
table {%
1 7631.25
2 14751.9
3 17491.38
4 20295.24
5 21654.25
6 23625.24
7 26244.47
8 29674
};
\addlegendentry{3-layer ansatz}
\addplot [semithick, forestgreen4416044]
table {%
1 6942.32
2 29938.02
3 33771.78
4 37392.88
5 40990.95
6 43853.04
7 49330.26
8 55166.16
};
\addlegendentry{4-layer ansatz}


\nextgroupplot[
legend cell align={left},
legend style={
  fill opacity=0.8,
  draw opacity=1,
  text opacity=1,
  at={(0.03,0.97)},
  anchor=north west,
  draw=lightgray204
},
tick align=outside,
tick pos=left,
x grid style={darkgray176},
xlabel={T},
xmin=0.65, xmax=8.35,
xtick distance={1},
xtick style={color=black},
y grid style={darkgray176},
ylabel={Circuit Runs},
ymin=-3973.5955, ymax=172969.0055,
ytick style={color=black}
]
\addplot [draw=steelblue31119180, fill=steelblue31119180, forget plot, mark=*, only marks]
table{%
x  y
1 4069.25
2 8624
3 13122.75
4 17369
5 21627.5
6 25341
7 29942.5
8 34912
};
\path [fill=steelblue31119180, fill opacity=0.3]
(axis cs:1,4820.62166236158)
--(axis cs:1,3317.87833763842)
--(axis cs:2,6808.44774242106)
--(axis cs:3,11437.2495179473)
--(axis cs:4,15041.4068224881)
--(axis cs:5,18581.3651540025)
--(axis cs:6,21444.9250520556)
--(axis cs:7,25130.9482752443)
--(axis cs:8,29366.8327347139)
--(axis cs:8,40457.1672652861)
--(axis cs:8,40457.1672652861)
--(axis cs:7,34754.0517247557)
--(axis cs:6,29237.0749479444)
--(axis cs:5,24673.6348459975)
--(axis cs:4,19696.5931775119)
--(axis cs:3,14808.2504820527)
--(axis cs:2,10439.5522575789)
--(axis cs:1,4820.62166236158)
--cycle;

\addplot [draw=darkorange25512714, fill=darkorange25512714, forget plot, mark=*, only marks]
table{%
x  y
1 8360.52
2 23474.28
3 35665.41
4 46312.16
5 55574
6 68023.02
7 78883.63
8 85807.44
};
\path [fill=darkorange25512714, fill opacity=0.3]
(axis cs:1,9895.31282510702)
--(axis cs:1,6825.72717489298)
--(axis cs:2,18325.9473114493)
--(axis cs:3,28695.4717172747)
--(axis cs:4,36822.3219841011)
--(axis cs:5,47552.1943522671)
--(axis cs:6,57554.2922818864)
--(axis cs:7,66559.7279528803)
--(axis cs:8,73137.1497138226)
--(axis cs:8,98477.7302861773)
--(axis cs:8,98477.7302861773)
--(axis cs:7,91207.5320471197)
--(axis cs:6,78491.7477181136)
--(axis cs:5,63595.8056477329)
--(axis cs:4,55801.9980158989)
--(axis cs:3,42635.3482827253)
--(axis cs:2,28622.6126885507)
--(axis cs:1,9895.31282510702)
--cycle;

\addplot [draw=forestgreen4416044, fill=forestgreen4416044, forget plot, mark=*, only marks]
table{%
x  y
1 10297.84
2 49141.12
3 71447.88
4 91424.2
5 104901.65
6 127272.6
7 150552.99
8 164926.16
};
\path [fill=forestgreen4416044, fill opacity=0.3]
(axis cs:1,11858.9556997481)
--(axis cs:1,8736.7243002519)
--(axis cs:2,36249.6686623111)
--(axis cs:3,58060.1694416558)
--(axis cs:4,73922.6785461378)
--(axis cs:5,86506.0798804033)
--(axis cs:6,104768.498717789)
--(axis cs:7,124099.503309981)
--(axis cs:8,133611.598769033)
--(axis cs:8,196240.721230967)
--(axis cs:8,196240.721230967)
--(axis cs:7,177006.476690019)
--(axis cs:6,149776.701282211)
--(axis cs:5,123297.220119597)
--(axis cs:4,108925.721453862)
--(axis cs:3,84835.5905583442)
--(axis cs:2,62032.5713376889)
--(axis cs:1,11858.9556997481)
--cycle;

\addplot [semithick, steelblue31119180]
table {%
1 4069.25
2 8624
3 13122.75
4 17369
5 21627.5
6 25341
7 29942.5
8 34912
};
\addlegendentry{2-layer ansatz}
\addplot [semithick, darkorange25512714]
table {%
1 8360.52
2 23474.28
3 35665.41
4 46312.16
5 55574
6 68023.02
7 78883.63
8 85807.44
};
\addlegendentry{3-layer ansatz}
\addplot [semithick, forestgreen4416044]
table {%
1 10297.84
2 49141.12
3 71447.88
4 91424.2
5 104901.65
6 127272.6
7 150552.99
8 164926.16
};
\addlegendentry{4-layer ansatz}
\end{groupplot}

\node[text width=6cm,align=center,anchor=north] at ($(group c1r1.south)-(0,1cm)$) {(a)};
\node[text width=6cm,align=center,anchor=north] at ($(group c2r1.south)-(0,1cm)$) {(d)};
\node[text width=6cm,align=center,anchor=north] at ($(group c1r2.south)-(0,1cm)$) {(b)};
\node[text width=6cm,align=center,anchor=north] at ($(group c2r2.south)-(0,1cm)$) {(e)};
\node[text width=6cm,align=center,anchor=north] at ($(group c1r3.south)-(0,1cm)$) {(c)};
\node[text width=6cm,align=center,anchor=north] at ($(group c2r3.south)-(0,1cm)$) {(f)};

\end{tikzpicture}
    \caption{The graphs depict the number of quantum circuits run when simulating the algorithm for generated matrices of various sizes. A solid line represents the number of quantum circuit evaluations averaged over all $100$ runs of the algorithm, and the shaded region around it represents $\pm 1$ standard deviation around the mean. (a), (b), and (c) are for $2\times2$, $4\times4$, and $8\times8$ matrices respectively using the original objective function. (d), (e), (f) are for $2\times2$, $4\times4$, and $8\times8$ matrices respectively using the modified objective function.}
    \label{fig:rand_evals}
\end{figure}
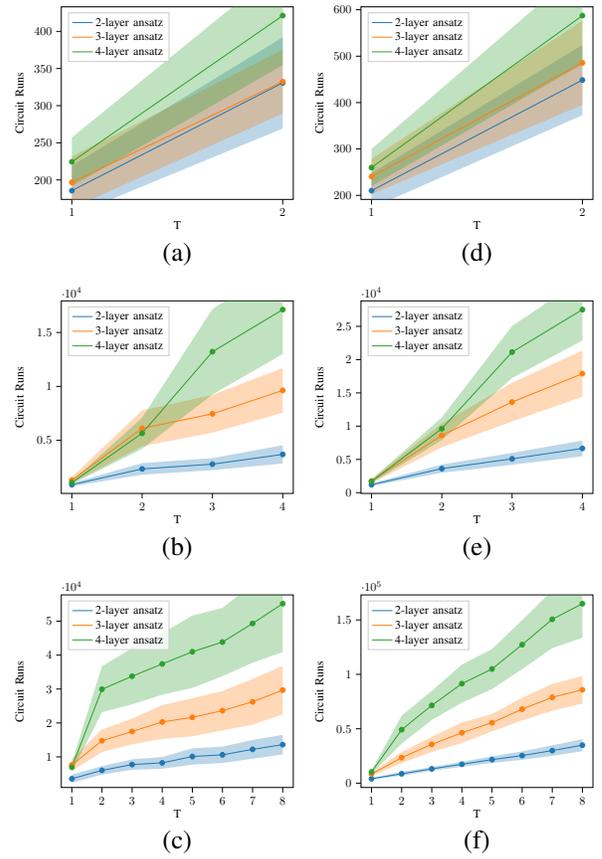
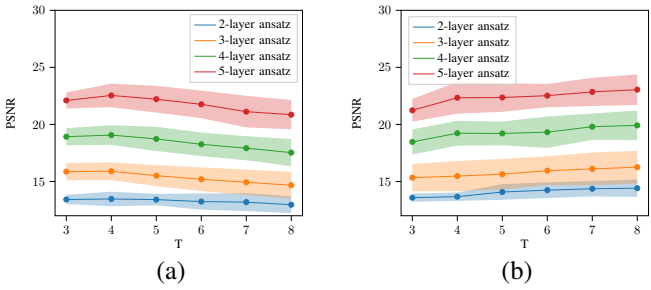
\begin{figure}[tb]
    \centering
    \hspace*{-2.3cm}
\begin{minipage}[b]{.44\columnwidth}
\begin{tikzpicture}[scale=0.48]
\definecolor{crimson2143940}{RGB}{214,39,40}
\definecolor{darkgray176}{RGB}{176,176,176}
\definecolor{darkorange25512714}{RGB}{255,127,14}
\definecolor{forestgreen4416044}{RGB}{44,160,44}
\definecolor{lightgray204}{RGB}{204,204,204}
\definecolor{steelblue31119180}{RGB}{31,119,180}

\begin{axis}[
legend cell align={left},
legend style={fill opacity=0.8, draw opacity=1, text opacity=1, draw=lightgray204},
tick align=outside,
tick pos=left,
x grid style={darkgray176},
xlabel={T},
xmin=2.75, xmax=8.25,
xtick style={color=black},
y grid style={darkgray176},
ylabel={PSNR},
ymin=12, ymax=30,
ytick style={color=black}
]
\addplot [draw=steelblue31119180, fill=steelblue31119180, forget plot, mark=*, only marks]
table{%
x  y
3 13.4218706980645
4 13.46791206909
5 13.4125379477302
6 13.2404833923436
7 13.1905038197423
8 12.9642091610246
};
\path [fill=steelblue31119180, fill opacity=0.3]
(axis cs:3,13.8275455603647)
--(axis cs:3,13.0161958357644)
--(axis cs:4,12.8295802412796)
--(axis cs:5,12.9058542084116)
--(axis cs:6,12.5298868417851)
--(axis cs:7,12.405504163703)
--(axis cs:8,12.2307051552503)
--(axis cs:8,13.6977131667989)
--(axis cs:8,13.6977131667989)
--(axis cs:7,13.9755034757817)
--(axis cs:6,13.9510799429021)
--(axis cs:5,13.9192216870488)
--(axis cs:4,14.1062438969004)
--(axis cs:3,13.8275455603647)
--cycle;

\addplot [draw=darkorange25512714, fill=darkorange25512714, forget plot, mark=*, only marks]
table{%
x  y
3 15.8636626542238
4 15.8986310253319
5 15.5111122907037
6 15.1982709771093
7 14.9300574955452
8 14.6731446047653
};
\path [fill=darkorange25512714, fill opacity=0.3]
(axis cs:3,16.6198514995481)
--(axis cs:3,15.1074738088996)
--(axis cs:4,15.122313472552)
--(axis cs:5,14.5860000907887)
--(axis cs:6,14.1598505997649)
--(axis cs:7,13.8193559417767)
--(axis cs:8,13.5007061192952)
--(axis cs:8,15.8455830902353)
--(axis cs:8,15.8455830902353)
--(axis cs:7,16.0407590493137)
--(axis cs:6,16.2366913544536)
--(axis cs:5,16.4362244906186)
--(axis cs:4,16.6749485781119)
--(axis cs:3,16.6198514995481)
--cycle;

\addplot [draw=forestgreen4416044, fill=forestgreen4416044, forget plot, mark=*, only marks]
table{%
x  y
3 18.9275576760672
4 19.0658808775904
5 18.7312225089045
6 18.2581677361003
7 17.917394883542
8 17.5306159981625
};
\path [fill=forestgreen4416044, fill opacity=0.3]
(axis cs:3,19.6811437791423)
--(axis cs:3,18.1739715729922)
--(axis cs:4,18.2026349578432)
--(axis cs:5,17.669062201455)
--(axis cs:6,17.2198621807653)
--(axis cs:7,16.8619149016841)
--(axis cs:8,16.3409273628131)
--(axis cs:8,18.7203046335119)
--(axis cs:8,18.7203046335119)
--(axis cs:7,18.9728748653998)
--(axis cs:6,19.2964732914353)
--(axis cs:5,19.793382816354)
--(axis cs:4,19.9291267973376)
--(axis cs:3,19.6811437791423)
--cycle;

\addplot [draw=crimson2143940, fill=crimson2143940, forget plot, mark=*, only marks]
table{%
x  y
3 22.1018550652044
4 22.5408534792539
5 22.2120044152142
6 21.7602563337473
7 21.1206921348354
8 20.8559543068225
};
\path [fill=crimson2143940, fill opacity=0.3]
(axis cs:3,22.8066326538389)
--(axis cs:3,21.3970774765698)
--(axis cs:4,21.5044418813226)
--(axis cs:5,21.0388236085338)
--(axis cs:6,20.5516103884966)
--(axis cs:7,19.7357298828468)
--(axis cs:8,19.5739720320465)
--(axis cs:8,22.1379365815986)
--(axis cs:8,22.1379365815986)
--(axis cs:7,22.5056543868239)
--(axis cs:6,22.968902278998)
--(axis cs:5,23.3851852218947)
--(axis cs:4,23.5772650771852)
--(axis cs:3,22.8066326538389)
--cycle;

\addplot [semithick, steelblue31119180]
table {%
3 13.4218706980645
4 13.46791206909
5 13.4125379477302
6 13.2404833923436
7 13.1905038197423
8 12.9642091610246
};
\addlegendentry{2-layer ansatz}
\addplot [semithick, darkorange25512714]
table {%
3 15.8636626542238
4 15.8986310253319
5 15.5111122907037
6 15.1982709771093
7 14.9300574955452
8 14.6731446047653
};
\addlegendentry{3-layer ansatz}
\addplot [semithick, forestgreen4416044]
table {%
3 18.9275576760672
4 19.0658808775904
5 18.7312225089045
6 18.2581677361003
7 17.917394883542
8 17.5306159981625
};
\addlegendentry{4-layer ansatz}
\addplot [semithick, crimson2143940]
table {%
3 22.1018550652044
4 22.5408534792539
5 22.2120044152142
6 21.7602563337473
7 21.1206921348354
8 20.8559543068225
};
\addlegendentry{5-layer ansatz}
\end{axis}
\node[text width=6cm,align=center,anchor=north] at ($(group c1r1.south)-(0,1cm)$) {(a)};
\end{tikzpicture}
\end{minipage}\qquad
\begin{minipage}[b]{.44\columnwidth}
\begin{tikzpicture}[scale=0.48]

\definecolor{crimson2143940}{RGB}{214,39,40}
\definecolor{darkgray176}{RGB}{176,176,176}
\definecolor{darkorange25512714}{RGB}{255,127,14}
\definecolor{forestgreen4416044}{RGB}{44,160,44}
\definecolor{lightgray204}{RGB}{204,204,204}
\definecolor{steelblue31119180}{RGB}{31,119,180}

\begin{axis}[
legend cell align={left},
legend style={
  fill opacity=0.8,
  draw opacity=1,
  text opacity=1,
  at={(0.03,0.97)},
  anchor=north west,
  draw=lightgray204
},
tick align=outside,
tick pos=left,
x grid style={darkgray176},
xlabel={T},
xmin=2.75, xmax=8.25,
xtick style={color=black},
y grid style={darkgray176},
ylabel={PSNR},
ymin=12, ymax=30,
ytick style={color=black}
]
\addplot [draw=steelblue31119180, fill=steelblue31119180, forget plot, mark=*, only marks]
table{%
x  y
3 13.5814076568077
4 13.6702318561786
5 14.0786827979969
6 14.2381364232662
7 14.3633427244407
8 14.4109336075703
};
\path [fill=steelblue31119180, fill opacity=0.3]
(axis cs:3,13.9466858119579)
--(axis cs:3,13.2161295016575)
--(axis cs:4,13.3033915010741)
--(axis cs:5,13.3915476268391)
--(axis cs:6,13.5508629949444)
--(axis cs:7,13.6857824664072)
--(axis cs:8,13.6437753956386)
--(axis cs:8,15.1780918195019)
--(axis cs:8,15.1780918195019)
--(axis cs:7,15.0409029824742)
--(axis cs:6,14.9254098515881)
--(axis cs:5,14.7658179691547)
--(axis cs:4,14.0370722112831)
--(axis cs:3,13.9466858119579)
--cycle;

\addplot [draw=darkorange25512714, fill=darkorange25512714, forget plot, mark=*, only marks]
table{%
x  y
3 15.3377531305212
4 15.4737060875306
5 15.637427392962
6 15.9481122602842
7 16.1038526678147
8 16.2530935984581
};
\path [fill=darkorange25512714, fill opacity=0.3]
(axis cs:3,16.5323022758431)
--(axis cs:3,14.1432039851992)
--(axis cs:4,14.1530570179662)
--(axis cs:5,14.2967725263862)
--(axis cs:6,14.685181050268)
--(axis cs:7,14.6707191511208)
--(axis cs:8,14.8178473017619)
--(axis cs:8,17.6883398951542)
--(axis cs:8,17.6883398951542)
--(axis cs:7,17.5369861845085)
--(axis cs:6,17.2110434703004)
--(axis cs:5,16.9780822595378)
--(axis cs:4,16.794355157095)
--(axis cs:3,16.5323022758431)
--cycle;

\addplot [draw=forestgreen4416044, fill=forestgreen4416044, forget plot, mark=*, only marks]
table{%
x  y
3 18.4687193523698
4 19.2299410911622
5 19.2108968533265
6 19.3195606901851
7 19.8013116236493
8 19.9161216871812
};
\path [fill=forestgreen4416044, fill opacity=0.3]
(axis cs:3,19.5689955388462)
--(axis cs:3,17.3684431658935)
--(axis cs:4,18.135300053947)
--(axis cs:5,18.1688832604213)
--(axis cs:6,17.9404776814581)
--(axis cs:7,18.6444251948081)
--(axis cs:8,18.6161448586068)
--(axis cs:8,21.2160985157557)
--(axis cs:8,21.2160985157557)
--(axis cs:7,20.9581980524904)
--(axis cs:6,20.6986436989121)
--(axis cs:5,20.2529104462317)
--(axis cs:4,20.3245821283775)
--(axis cs:3,19.5689955388462)
--cycle;

\addplot [draw=crimson2143940, fill=crimson2143940, forget plot, mark=*, only marks]
table{%
x  y
3 21.2345898444885
4 22.3386256060609
5 22.3592769759593
6 22.5213219295016
7 22.8505465234783
8 23.0478402323922
};
\path [fill=crimson2143940, fill opacity=0.3]
(axis cs:3,22.2274478721811)
--(axis cs:3,20.2417318167959)
--(axis cs:4,20.9455115803933)
--(axis cs:5,21.1058142024598)
--(axis cs:6,21.4978475157353)
--(axis cs:7,21.6211412582589)
--(axis cs:8,21.7051170027051)
--(axis cs:8,24.3905634620793)
--(axis cs:8,24.3905634620793)
--(axis cs:7,24.0799517886976)
--(axis cs:6,23.544796343268)
--(axis cs:5,23.6127397494588)
--(axis cs:4,23.7317396317284)
--(axis cs:3,22.2274478721811)
--cycle;

\addplot [semithick, steelblue31119180]
table {%
3 13.5814076568077
4 13.6702318561786
5 14.0786827979969
6 14.2381364232662
7 14.3633427244407
8 14.4109336075703
};
\addlegendentry{2-layer ansatz}
\addplot [semithick, darkorange25512714]
table {%
3 15.3377531305212
4 15.4737060875306
5 15.637427392962
6 15.9481122602842
7 16.1038526678147
8 16.2530935984581
};
\addlegendentry{3-layer ansatz}
\addplot [semithick, forestgreen4416044]
table {%
3 18.4687193523698
4 19.2299410911622
5 19.2108968533265
6 19.3195606901851
7 19.8013116236493
8 19.9161216871812
};
\addlegendentry{4-layer ansatz}
\addplot [semithick, crimson2143940]
table {%
3 21.2345898444885
4 22.3386256060609
5 22.3592769759593
6 22.5213219295016
7 22.8505465234783
8 23.0478402323922
};
\addlegendentry{5-layer ansatz}
\end{axis}
\node[text width=6cm,align=center,anchor=north] at ($(group c1r1.south)-(0,1cm)$) {(b)};
\end{tikzpicture}
\end{minipage}
    \caption{The graphs depict the PSNR of the reconstructed images from simulating the algorithm on the MNIST dataset. A solid line represents the PSNR averaged over all $200$ runs of the algorithm, and the shaded region around it represents $\pm 1$ standard deviation around the mean. (a) depicts the results when using the original objective function while (b) depicts the results using the modified objective function.}
    \label{fig:img_psnr}
\end{figure}

\section{Benchmarking}\label{sec:bench}

In this section we run simulations to benchmark this algorithm using both the objective function from \cite{wang} and our new objective function. Table \ref{table:orig} and Table \ref{table:new} show the circuit depths and gate counts for the original and modified objective functions respectively. We plot the average accuracy of the solutions returned by the algorithm using appropriate metrics and we plot the average number of circuit evaluations (not taking shots into account; we assume infinite shots) required for one run of this algorithm. The quantum circuit simulations are run using PennyLane \cite{pennylane}. We benchmarked the algorithm for randomly generated matices of various sizes and then we did the same for the first $20$ $8\times8$ images from the MNIST Handwritten Digit dataset \cite{mnist} loaded using \texttt{scikit-learn} \cite{scikit-learn}. All simulations make use of our new method for evaluating expectation values. This decision should not affect any of the results or graphs except Table \ref{table:orig} and Table \ref{table:new}.

\subsection{Mean Square Error with randomly generated matrices}

In this subsection we measure the performance of the algorithm on randomly generated $2^n\times2^n$ square matrices for $n=1,2,3$. The matrices are generated such that each matrix entry is sampled from a uniform distribution in $[0,1)$. For each $1\leq T\leq 2^n$, and for each of $10$ randomly generated matrices, we run simulations with $10$ randomly generated initial parameters. For each run of the algorithm, we use the returned solution to reconstruct the input matrix and we evaluate the MSE of the reconstructed matrix as described in Section \ref{sec:obj}. In Fig. \ref{fig:rand_mse} we plot the average MSE of these $100$ runs against $T$ using the original and modified objective functions. As we can see, the issue seen in Fig. \ref{fig:orig_fun_eg} takes shape when we use $8\times8$ matrices when using the original objective function.

Fig. \ref{fig:rand_evals} plots the average number of circuit runs for the same simulations for the original and modified objective functions.

\subsection{Image compression}

An example of a real application that could make use of this algorithm is image compression. We ran simulations for $20$ $8\times8$ images from the MNIST Handwritten Digit dataset for $10$ random initial parameters each and measured the averaged PSNR of the reconstructed images. The average PSNR is plotted for various ansatz sizes against $3\leq T\leq8$ in Fig. \ref{fig:img_psnr} for the original and modified objective functions.

\begin{figure}[tb]
    \centering
    \hspace*{-2.3cm}
\begin{minipage}[b]{.44\columnwidth}
\begin{tikzpicture}[scale=0.48]

\definecolor{crimson2143940}{RGB}{214,39,40}
\definecolor{darkgray176}{RGB}{176,176,176}
\definecolor{darkorange25512714}{RGB}{255,127,14}
\definecolor{forestgreen4416044}{RGB}{44,160,44}
\definecolor{lightgray204}{RGB}{204,204,204}
\definecolor{steelblue31119180}{RGB}{31,119,180}

\begin{axis}[
legend cell align={left},
legend style={
  fill opacity=0.8,
  draw opacity=1,
  text opacity=1,
  at={(0.03,0.97)},
  anchor=north west,
  draw=lightgray204
},
tick align=outside,
tick pos=left,
x grid style={darkgray176},
xlabel={T},
xmin=2.75, xmax=8.25,
xtick style={color=black},
y grid style={darkgray176},
ylabel={Circuit Runs},
ymin=330.547999999999, ymax=143921.492,
ytick style={color=black}
]
\addplot [draw=steelblue31119180, fill=steelblue31119180, forget plot, mark=*, only marks]
table{%
x  y
3 7498.125
4 8238.5
5 9118.125
6 10142.25
7 11100.25
8 13541
};
\path [fill=steelblue31119180, fill opacity=0.3]
(axis cs:3,8783.27905204979)
--(axis cs:3,6212.97094795021)
--(axis cs:4,6647.8161847809)
--(axis cs:5,7027.20900754597)
--(axis cs:6,8052.33762145395)
--(axis cs:7,8554.8029616588)
--(axis cs:8,10182.2055734237)
--(axis cs:8,16899.7944265763)
--(axis cs:8,16899.7944265763)
--(axis cs:7,13645.6970383412)
--(axis cs:6,12232.162378546)
--(axis cs:5,11209.040992454)
--(axis cs:4,9829.1838152191)
--(axis cs:3,8783.27905204979)
--cycle;

\addplot [draw=darkorange25512714, fill=darkorange25512714, forget plot, mark=*, only marks]
table{%
x  y
3 16897.53
4 19877.88
5 22668.975
6 25361.28
7 28021.21
8 32059.76
};
\path [fill=darkorange25512714, fill opacity=0.3]
(axis cs:3,20654.0133786402)
--(axis cs:3,13141.0466213598)
--(axis cs:4,14446.7215983328)
--(axis cs:5,17072.5751672951)
--(axis cs:6,18780.314755904)
--(axis cs:7,21136.9237399727)
--(axis cs:8,23495.11367969)
--(axis cs:8,40624.40632031)
--(axis cs:8,40624.40632031)
--(axis cs:7,34905.4962600273)
--(axis cs:6,31942.245244096)
--(axis cs:5,28265.374832705)
--(axis cs:4,25309.0384016672)
--(axis cs:3,20654.0133786402)
--cycle;

\addplot [draw=forestgreen4416044, fill=forestgreen4416044, forget plot, mark=*, only marks]
table{%
x  y
3 33673.29
4 40617.08
5 47898.725
6 53205.18
7 59853.5
8 65710.96
};
\path [fill=forestgreen4416044, fill opacity=0.3]
(axis cs:3,41093.1038967228)
--(axis cs:3,26253.4761032773)
--(axis cs:4,29233.005847044)
--(axis cs:5,35695.5494660457)
--(axis cs:6,38535.7849192204)
--(axis cs:7,42354.4850831254)
--(axis cs:8,49680.0194721333)
--(axis cs:8,81741.9005278667)
--(axis cs:8,81741.9005278667)
--(axis cs:7,77352.5149168746)
--(axis cs:6,67874.5750807796)
--(axis cs:5,60101.9005339543)
--(axis cs:4,52001.154152956)
--(axis cs:3,41093.1038967228)
--cycle;

\addplot [draw=crimson2143940, fill=crimson2143940, forget plot, mark=*, only marks]
table{%
x  y
3 63820.335
4 74865.3
5 86775.55
6 95073.99
7 108052.35
8 123049.2
};
\path [fill=crimson2143940, fill opacity=0.3]
(axis cs:3,78647.8454896776)
--(axis cs:3,48992.8245103224)
--(axis cs:4,56035.4791556053)
--(axis cs:5,59728.2796631434)
--(axis cs:6,69882.0276121669)
--(axis cs:7,77020.6174419877)
--(axis cs:8,91372.169350269)
--(axis cs:8,154726.230649731)
--(axis cs:8,154726.230649731)
--(axis cs:7,139084.082558012)
--(axis cs:6,120265.952387833)
--(axis cs:5,113822.820336857)
--(axis cs:4,93695.1208443947)
--(axis cs:3,78647.8454896776)
--cycle;

\addplot [semithick, steelblue31119180]
table {%
3 7498.125
4 8238.5
5 9118.125
6 10142.25
7 11100.25
8 13541
};
\addlegendentry{2-layer ansatz}
\addplot [semithick, darkorange25512714]
table {%
3 16897.53
4 19877.88
5 22668.975
6 25361.28
7 28021.21
8 32059.76
};
\addlegendentry{3-layer ansatz}
\addplot [semithick, forestgreen4416044]
table {%
3 33673.29
4 40617.08
5 47898.725
6 53205.18
7 59853.5
8 65710.96
};
\addlegendentry{4-layer ansatz}
\addplot [semithick, crimson2143940]
table {%
3 63820.335
4 74865.3
5 86775.55
6 95073.99
7 108052.35
8 123049.2
};
\addlegendentry{5-layer ansatz}
\end{axis}

\node[text width=6cm,align=center,anchor=north] at ($(group c1r1.south)-(0,1cm)$) {(a)};

\end{tikzpicture}
\end{minipage}\qquad
\begin{minipage}[b]{.44\columnwidth}
\begin{tikzpicture}[scale=0.48]

\definecolor{crimson2143940}{RGB}{214,39,40}
\definecolor{darkgray176}{RGB}{176,176,176}
\definecolor{darkorange25512714}{RGB}{255,127,14}
\definecolor{forestgreen4416044}{RGB}{44,160,44}
\definecolor{lightgray204}{RGB}{204,204,204}
\definecolor{steelblue31119180}{RGB}{31,119,180}

\begin{axis}[
legend cell align={left},
legend style={
  fill opacity=0.8,
  draw opacity=1,
  text opacity=1,
  at={(0.03,0.97)},
  anchor=north west,
  draw=lightgray204
},
tick align=outside,
tick pos=left,
x grid style={darkgray176},
xlabel={T},
xmin=2.75, xmax=8.25,
xtick style={color=black},
y grid style={darkgray176},
ylabel={Circuit Runs},
ymin=319.547999999999, ymax=273921.492,
ytick style={color=black}
]
\addplot [draw=steelblue31119180, fill=steelblue31119180, forget plot, mark=*, only marks]
table{%
x  y
3 12756
4 16991
5 21334.375
6 25001.25
7 29093.75
8 33433
};
\path [fill=steelblue31119180, fill opacity=0.3]
(axis cs:3,14850.3634176045)
--(axis cs:3,10661.6365823955)
--(axis cs:4,14370.5706458674)
--(axis cs:5,18379.7850855618)
--(axis cs:6,21495.6646351857)
--(axis cs:7,24937.921863515)
--(axis cs:8,28813.2878877575)
--(axis cs:8,38052.7121122425)
--(axis cs:8,38052.7121122425)
--(axis cs:7,33249.578136485)
--(axis cs:6,28506.8353648143)
--(axis cs:5,24288.9649144382)
--(axis cs:4,19611.4293541326)
--(axis cs:3,14850.3634176045)
--cycle;

\addplot [draw=darkorange25512714, fill=darkorange25512714, forget plot, mark=*, only marks]
table{%
x  y
3 33806.16
4 44209.08
5 54854.35
6 65594.34
7 73482.185
8 85788.2
};
\path [fill=darkorange25512714, fill opacity=0.3]
(axis cs:3,39523.8873342649)
--(axis cs:3,28088.4326657351)
--(axis cs:4,36736.0735709917)
--(axis cs:5,45791.1529854251)
--(axis cs:6,55002.3340313461)
--(axis cs:7,62021.1394218906)
--(axis cs:8,72671.7873599524)
--(axis cs:8,98904.6126400476)
--(axis cs:8,98904.6126400476)
--(axis cs:7,84943.2305781094)
--(axis cs:6,76186.3459686539)
--(axis cs:5,63917.5470145749)
--(axis cs:4,51682.0864290083)
--(axis cs:3,39523.8873342649)
--cycle;

\addplot [draw=forestgreen4416044, fill=forestgreen4416044, forget plot, mark=*, only marks]
table{%
x  y
3 62940.255
4 85015
5 103779.55
6 123072.81
7 143476.9
8 165580.8
};
\path [fill=forestgreen4416044, fill opacity=0.3]
(axis cs:3,75341.9067547261)
--(axis cs:3,50538.6032452739)
--(axis cs:4,69403.7027373123)
--(axis cs:5,86731.1584996267)
--(axis cs:6,101703.071589481)
--(axis cs:7,119911.581807325)
--(axis cs:8,138584.809888282)
--(axis cs:8,192576.790111718)
--(axis cs:8,192576.790111718)
--(axis cs:7,167042.218192675)
--(axis cs:6,144442.548410519)
--(axis cs:5,120827.941500373)
--(axis cs:4,100626.297262688)
--(axis cs:3,75341.9067547261)
--cycle;

\addplot [draw=crimson2143940, fill=crimson2143940, forget plot, mark=*, only marks]
table{%
x  y
3 102831.36
4 136120.28
5 169471.725
6 198454.35
7 231476.7
8 261485.04
};
\path [fill=crimson2143940, fill opacity=0.3]
(axis cs:3,123081.028177504)
--(axis cs:3,82581.6918224965)
--(axis cs:4,108686.45274648)
--(axis cs:5,138481.427828748)
--(axis cs:6,165199.046402198)
--(axis cs:7,190111.856555125)
--(axis cs:8,213422.296815367)
--(axis cs:8,309547.783184633)
--(axis cs:8,309547.783184633)
--(axis cs:7,272841.543444875)
--(axis cs:6,231709.653597802)
--(axis cs:5,200462.022171252)
--(axis cs:4,163554.10725352)
--(axis cs:3,123081.028177504)
--cycle;

\addplot [semithick, steelblue31119180]
table {%
3 12756
4 16991
5 21334.375
6 25001.25
7 29093.75
8 33433
};
\addlegendentry{2-layer ansatz}
\addplot [semithick, darkorange25512714]
table {%
3 33806.16
4 44209.08
5 54854.35
6 65594.34
7 73482.185
8 85788.2
};
\addlegendentry{3-layer ansatz}
\addplot [semithick, forestgreen4416044]
table {%
3 62940.255
4 85015
5 103779.55
6 123072.81
7 143476.9
8 165580.8
};
\addlegendentry{4-layer ansatz}
\addplot [semithick, crimson2143940]
table {%
3 102831.36
4 136120.28
5 169471.725
6 198454.35
7 231476.7
8 261485.04
};
\addlegendentry{5-layer ansatz}
\end{axis}
\node[text width=6cm,align=center,anchor=north] at ($(group c1r1.south)-(0,1cm)$) {(b)};

\end{tikzpicture}
\end{minipage}
\caption{The graphs depict the number of quantum circuits run when simulating the algorithm for MNIST images. A solid line represents the number of quantum circuit evaluations averaged over all $200$ runs of the algorithm, and the shaded region around it represents $\pm 1$ standard deviation around the mean. (a) depicts the results when using the original objective function while (b) depicts the results using the modified objective function.}
\label{fig:img_evals}
\end{figure}
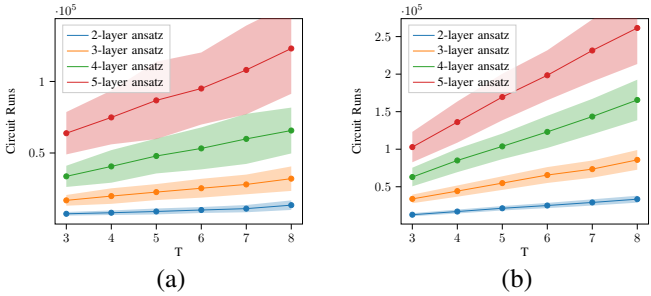

Fig. \ref{fig:img_evals} plots the average number of circuit runs for the same simulations for the original and new objective functions.

\section{Discussion and Future Works}\label{sec:discuss}
Our first modification to the algorithm put forward in \cite{wang} was to the objective function. This led to more consistent trends than seen with the original objective function. Further tests need to be done to make sure that this is the case for larger problems. While this hasn't been practically feasible using gate level quantum simulators like those in Qiskit \cite{Qiskit} and PennyLane, numerical studies could be done.\\
One disadvantage to our objective function however is that it seems to require more quantum circuit evaluations to get to a solution. Furthermore, an inconvenience with our objective function is that the values we get for $\Tilde{\sigma}_i$ may generally be complex values, but this is not too grave an issue. To get the SVD, we simply have to take out the complex phases from $\Tilde{\sigma}_i$ manually and absorb them into the corresponding left or right singular vectors.
Another smaller inconvenience is that the values we get are not received in descending order of magnitude like they are in \cite{wang}.

Moving on to our second modification, it is not obvious whether our new technique for evaluating expectation values generally hosts any advantage for this algorithm, but it could be useful in specific scenarios. For example, there could be cases where we may only have quantum access to the matrix $A$ that we want the SVD of, and in such a case this new approach could be useful, since it takes a quantum statevector encoding $A$ directly as input. Applications of this new block encoding method must also be explored for other problems beyond the scope of this variational quantum SVD algorithm. Further work also needs to be done in benchmarking this algorithm against the one proposed in \cite{qsvdecomposer}.
\bibliographystyle{IEEEtran}
\bibliography{IEEEabrv,citations}

\end{document}